\documentclass[12pt]{article}
\usepackage[left=1in,right=1in,top=1in,bottom=1in]{geometry}
\usepackage{graphicx}
\usepackage{float}
\usepackage{amsfonts}
\usepackage{amssymb}
\usepackage{amsmath}
\usepackage{relsize}

\def\gtwid{\mathrel{\raise.3ex\hbox{$>$\kern-.75em\lower1ex\hbox{$\sim$}}}}
\def\ltwid{\mathrel{\raise.3ex\hbox{$<$\kern-.75em\lower1ex\hbox{$\sim$}}}}
\def\square{\kern1pt\vbox{\hrule height 1.2pt\hbox{\vrule width 1.2pt\hskip 3pt
   \vbox{\vskip 6pt}\hskip 3pt\vrule width 0.6pt}\hrule height 0.6pt}\kern1pt}

\begin{document}

\begin{titlepage}

\begin{flushright}
UFIFT-QG-21-03
\end{flushright}

\vskip 4cm

\begin{center}
{\bf Gauge Independent Quantum Gravitational Corrections to Maxwell's Equation}
\end{center}

\vskip 1cm

\begin{center}
S. Katuwal$^{*}$ and R. P. Woodard$^{\dagger}$
\end{center}

\begin{center}
\it{Department of Physics, University of Florida, \\
Gainesville, FL 32611, UNITED STATES}
\end{center}

\vspace{1cm}

\begin{center}
ABSTRACT
\end{center}
We consider quantum gravitational corrections to Maxwell's
equations on flat space background. Although the vacuum polarization
is highly gauge dependent, we explicitly show that this gauge dependence
is canceled by contributions from the source which disturbs the effective 
field and the observer who measures it. Our final result is a gauge
independent, real and causal effective field equation that can be used in
the same way as the classical Maxwell equation. 

\begin{flushleft}
PACS numbers: 04.50.Kd, 95.35.+d, 98.62.-g
\end{flushleft}

\vspace{6cm}

\begin{flushleft}
$^{*}$ e-mail: sanjib.katuwal@ufl.edu \\
$^{\dagger}$ e-mail: woodard@phys.ufl.edu
\end{flushleft}

\end{titlepage}

\section{Introduction}

The greatest story ever told in physics is how a century of brilliant 
experimental extemporization culminated in the development of Maxwell's
equations. This was humanity's first relativistic, unified field theory
and it set the stage for the discoveries of general relativity and 
non-Abelian gauge theories. Electrodynamics is still one of the core 
subjects in the study of physics. Most western physicists recall the 
ingenuity and perseverance required of them as graduate students to 
solve Maxwell's equations in the wide variety of settings treated in 
the classic text by the late J. D. Jackson \cite{Jackson:1998nia}.

Quantum loop corrections to electrodynamics are small at low frequencies,
and those from quantum gravity are unobservable. One might therefore
expect that including these effects causes only a small change in 
electrodynamics. The math is simple enough: one first computes the 1PI 
(one-particle-irreducible) 2-photon function, $i[\mbox{}^{\mu} 
\Pi^{\nu}](x;x')$, known as the ``vacuum polarization''. Then Maxwell's 
equations are supplemented by the integral of the vacuum polarization 
contracted into the vector potential $A_{\nu}(x')$,
\begin{equation}
\partial_{\nu} F^{\nu\mu}(x) + \int \!\! d^4x' \Bigl[\mbox{}^{\mu} 
\Pi^{\nu}\Bigr](x;x') A_{\nu}(x') = J^{\mu}(x) \; , \label{QMax}
\end{equation}
where $F_{\mu\nu} \equiv \partial_{\mu} A_{\nu} - \partial_{\nu} A_{\mu}$
is the field strength tensor and $J^{\mu}(x)$ is the current density.
However, students of quantum field theory are strongly enjoined that
they cannot think of solving the quantum-corrected equation the same
as its classical analog; they must instead abandon the concept of local 
fields and infer physics entirely from scattering amplitudes. Although 
basing physics on scattering amplitudes is valid for most situations on 
flat space background, it does seem to be an over-reaction, and it is 
not even possible in cosmology. The purpose of this paper is to provide 
a version of the quantum-corrected field equation (\ref{QMax}) which can 
be solved as in classical electrodynamics.

Part of the reason for the curious dichotomy between classical and 
quantum is the prevalence of the ``in-out'' formalism so elegantly
summarized by the Feynman rules. The in-out vacuum polarization is 
neither real, nor is it causal in the sense of vanishing for points
${x'}^{\mu}$ outside the past light-cone of $x^{\mu}$. Those two 
properties are not errors; in-out amplitudes are precisely the right 
objects of study for computing scattering amplitudes. However, the 
absence of reality and causality is certainly problematic if one 
wishes to regard solutions to the quantum-corrected field equation
(\ref{QMax}) as electric and magnetic fields. 

Julian Schwinger long ago devised a method for computing true 
expectation values which is almost as simple to use as the Feynman 
rules \cite{Schwinger:1960qe}. When the vacuum polarization of the 
Schwinger-Keldysh formalism is employed in equation (\ref{QMax}) the 
effective field equations become manifestly real and causal 
\cite{Mahanthappa:1962ex,Bakshi:1962dv,Bakshi:1963bn,Keldysh:1964ud,
Chou:1984es,Jordan:1986ug,Calzetta:1986ey,Ford:2004wc}. However, 
there is still an obstacle: the propagators of vector and tensor 
fields require gauge fixing, and loop corrections involving these 
propagators cause the vacuum polarization to depend on the choice 
of gauge. For example, single graviton loop corrections to the vacuum 
polarization on a $D$-dimensional flat space background ($g_{\mu\nu} 
\equiv \eta_{\mu\nu} + \kappa h_{\mu\nu}$ with $\kappa^2 = 16 \pi G$) 
with the most general, Poincar\'e invariant gauge fixing functional,
\begin{equation}
\mathcal{L}_{GF} = -\frac1{2 a} \eta^{\mu\nu} F_{\mu} F_{\nu} \qquad ,
\qquad F_{\mu} = \eta^{\rho\sigma} \Bigl(h_{\mu \rho , \sigma} -
\frac{b}{2} h_{\rho\sigma , \mu}\Bigr) \; , \label{gauge}
\end{equation}
result in a primitive vacuum polarization of the form \cite{Leonard:2012fs},
\begin{equation}
i\Bigl[\mbox{}^{\mu} \Pi^{\nu}\Bigr](x;x') = -\frac{\kappa^2 
\mathcal{C}_0(D,a,b) (D\!-\!2) \Gamma^2(\frac{D}2 \!-\! 1)}{32 (D\!-\!1) 
\pi^D} \Bigl( \eta^{\mu\nu} \partial^2 \!-\! \partial^{\mu} \partial^{\nu}
\Bigr) \frac{1}{(x \!-\! x')^{2D-2}} \; , \label{vacpol}
\end{equation}
where the gauge dependent multiplicative factor is,
\begin{eqnarray}
\lefteqn{\mathcal{C}_0(D,a,b) = D (D \!-\! 2) (D\!-\!3) + \frac{(D\!-\!1) 
(D\!-\!2)^2 [(D\!-\!2) (a \!-\! 1) \!-\! D (b \!-\! 1)^2]}{2 (b \!-\! 2)^2} }
\nonumber \\
& & \hspace{5.5cm} + (D\!-\!1) (D\!-\!2)^2 (D\!-\! 4) \Bigl[ -
\frac{(a \!-\!1)}{2} \!+\! \frac{2}{D \!-\! 2} \Bigl( \frac{b \!-\! 1}{
b \!-\! 2}\Bigr)\Bigr] \; . \label{C0def} \qquad 
\end{eqnarray}
Although the tensor structure and spacetime dependence of (\ref{vacpol})
is universal, the multiplicative factor $\mathcal{C}(D,a,b)$ can be made
to range from $-\infty$ to $+\infty$ by adjusting the gauge parameters 
$a$ and $b$ \cite{Leonard:2012fs}.

John Donoghue has shown how to use general relativity as a low energy
effective field theory to reliably compute quantum gravitational 
corrections to the long-range potentials induced by the exchange of 
massless particles such as photons and gravitons \cite{Donoghue:1993eb,
Donoghue:1994dn}. His technique is to compute the scattering amplitude 
between two massive particles which interact with the massless field,
and then use inverse scattering theory to infer the exchange potential.
In this way one can derive gauge independent, single graviton loop
corrections to the Newtonian potential \cite{Bjerrum-Bohr:2002fji,
Bjerrum-Bohr:2002gqz} and to the Coulomb potential 
\cite{Bjerrum-Bohr:2002aqa}. 

It has recently been noted that Donoghue's S-matrix technique can be 
short-circuited to produce gauge independent effective field equations 
directly, without passing through the intermediate stages of computing 
scattering amplitudes and solving the inverse scattering problem 
\cite{Miao:2017feh}. The key is applying position space versions of a 
series of identities derived by Donoghue and collaborators for the 
purpose of isolating the nonlocal and nonanalytic parts of scattering 
amplitudes which correct long-range potentials \cite{Donoghue:1994dn,
Donoghue:1996mt}. These identities degenerate the massive propagators 
of the particles being scattered to delta functions, thus casting the
important parts of higher-point contributions to 2-particle scattering 
in a form that can be regarded as corrections to the 1PI 2-point 
function of the massless field. In this picture the gauge dependence 
of the original effective field equation derives from having omitted 
to include quantum gravitational interactions with the source which 
disturbs the effective field and from the observer who measures it; 
and the corrections to the 1PI 2-point function repair this omission. 
The new technique has already been implemented at one loop order for 
quantum gravitational corrections to a massless scalar on flat space
background, and its independence of the gauge parameters $a$ and $b$ 
explicitly demonstrated \cite{Miao:2017feh}. In this paper we do the
same for quantum gravitational corrections to electrodynamics, which
is a realistic system and one involving vector fields.

This paper closely follows the analysis of Bjerrum-Bohr 
\cite{Bjerrum-Bohr:2002aqa} who applied Donoghue's technique to 
include one graviton loop corrections to electrodynamics on flat space 
background. Section 2 goes through the position-space version of each 
of the same diagrams he considered, including first order perturbations
of the gauge parameters (\ref{gauge}),
\begin{equation}
\left\{\begin{matrix}
a \equiv 1 + \delta a \cr b \equiv 1 + \delta b
\end{matrix} \right\} \qquad \Longrightarrow \qquad \mathcal{C}_0(4,a,b) 
= 8 + 12 \!\cdot\! \delta a + 0 \!\cdot\! \delta b + O(\delta^2) \; . 
\label{deltagauge}
\end{equation}
In each case we show how the Donoghue identities allow one to regard 
the diagram as a correction to the vacuum polarization. Of course the 
gauge dependence cancels when everything is summed up, and the result 
has the same form (\ref{vacpol}), but with the constant 
$\mathcal{C}_0(4,a,b)$ replaced by the gauge independent number $+40$.
Our conclusions comprise section 3. Three appendices give, respectively, 
the vertices, the propagators and the Donoghue identities, including
the new one we required for certain of the $\delta b$ contributions.

\section{Including the Source and the Observer}

In this section, we use the scattering of a pair of massive, charged
scalars to provide the source which disturbs the effective field and 
the observer who measures this disturbance. The Lagrangian that describes
the scattering is,
\begin{equation}\label{lagrangian}
\mathcal{L} = \left[\frac{R}{16\pi G}-\frac{1}{4} g^{\alpha\mu} g^{\beta\nu} 
F_{\alpha\nu} F_{\mu\beta} - (D_\mu\phi)^* g^{\mu\nu} (D_\nu\phi) + m^2 \phi^* 
\phi \right] \sqrt{-g} \; ,
\end{equation}
where, $D_{\mu} = \nabla_{\mu} - i e A_{\mu}$ and $\nabla_\mu$ denotes the 
metric-compatible covariant derivative. (When acting on a scalar the covariant
derivative degenerates to the partial derivative, $D_{\mu} \phi = 
\partial_{\mu} \phi - i e A_{\mu} \phi$.) Unless otherwise stated, we work with 
the usual $c = \hbar =1$ convention of particle physics, however, we employ a
spacelike metric. General relativity plus SQED (Scalar Quantum Electrodynamics) is treated
as a low energy effective field theory in the sense of Donoghue 
\cite{Donoghue:1993eb,Donoghue:1994dn}. The perturbation is around flat space 
with the following definitions of the graviton field $h_{\mu\nu}$ and the loop 
counting parameter $\kappa^2$,
\begin{equation}
g_{\mu\nu}(x) \equiv \eta_{\mu\nu} + \kappa h_{\mu\nu} , \quad 
\kappa^2 \equiv 16 \pi G \; .
\end{equation}
The vertices we require are listed in Appendix A, and the various propagators 
are given in Appendix B.

\setcounter{figure}{-1}
\begin{figure}[H]
\centering
\includegraphics[width=0.95\textwidth]{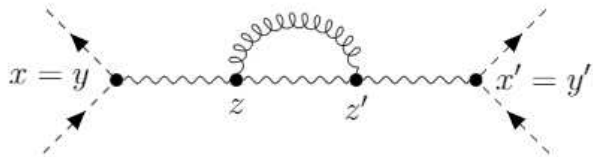}
\caption{\footnotesize This diagram shows how the vacuum polarization contributes to 
the amputated 4-scalar vertex function. Dashed lines represent massive scalars, wavy 
lines represent photons and curly lines represent gravitons. These graphs have the 
same topology as Bjerrum-Bohr's Diagram 8 \cite{Bjerrum-Bohr:2002aqa}.}
\label{fig:feyn_so}
\end{figure}

Our procedure for purging gauge dependence from the one loop vacuum polarization
is to write down position space representations for each of the order $e^2 \kappa^2$ 
contributions to the amputated 4-scalar function. Any external derivatives are assumed 
to act on the external scalar wave functions appropriate to 2-particle scattering. By 
exploiting the various Donoghue Identities of Appendix C to degenerate the (internal)
massive scalar propagators to Dirac delta functions, we reduce each contribution to
a form that can be interpreted as a correction to $i[\mbox{}^{\mu} \Pi^{\nu}](x;x')$.

We begin by considering the contribution of the original, gauge dependent vacuum 
polarization to the amputated 4-scalar function as shown in Figure \ref{fig:feyn_so}.
The expression for this diagram is
\begin{equation}\label{eq:so}
    \begin{split}
        iV_0(x;x') = & e (\partial_{x}\downarrow - \partial_{x} \uparrow)^{\alpha}
        \times e(\partial_{x'} \downarrow - \partial_{x'} \uparrow)^{\beta} \\
        & \times \int d^Dz\ i[_\alpha\Delta_\mu](x;z) \int d^Dz'\ i[_\beta\Delta_\nu](x';z')
        \times i[^\mu\Pi^\nu](z;z')
    \end{split}
\end{equation}
where the vacuum polarization was given in (\ref{vacpol}) and external derivatives with 
an up (down) arrow act on upper (lower) scalar wave functions at that vertex. First
note that Poincar\'e invariance and partial integration allows us to act all longitudinal
parts on the external legs, where (by current conservation) they vanish due to the 
on-shell condition,
\begin{equation}
(\partial \downarrow - \partial \uparrow)_{\mu} (\partial \downarrow + \partial 
\uparrow)^{\mu} = (\partial^2 \downarrow - m^2) - (\partial^2 \uparrow - m^2) \; .
\end{equation}
We can also use the relation,
\begin{equation}
\frac1{\Delta x^{2D-2}} = \frac{\partial^2}{2 (D\!-\! 2)^2} \frac1{\Delta x^{2D-4}}
= \frac{\partial^2}{2 (D\!-\!2)^2} \Biggl[ \frac{4 \pi^{\frac{D}2}}{\Gamma(
\frac{D}2 \!-\!1)} i\Delta(x;x')\Biggr]^2 \; ,
\end{equation}
to attain the form,
\begin{equation}\label{eq:vp_explicit}
\begin{split}
    iV_0(x;x') = & - \frac{e^2 \kappa^2 \mathcal{C}_0(D,a,b)}{4 (D\!-\! 1)(D\!-\!2)}
    \times (\partial_{x} \downarrow - \partial_{x} \uparrow)_{\mu}
    (\partial_{x'} \downarrow - \partial_{x'} \uparrow)^{\mu} \\
    & \times \int d^Dz\ i\Delta(x;z) \int d^Dz'\ i\Delta(x';z') \, 
    \partial_{z}^2 \partial_{z'}^2 \Bigl[ i\Delta(z;z')\Bigr]^2 \; .
\end{split}
\end{equation}
The final step is to partially integrate the factors of $\partial_{z}^2$ and 
$\partial_{z'}^2$ to act on the massless propagators, and use the delta functions
that result from the propagator equation (\ref{eq:equation_massless}) to eliminate
the integrations over $z^{\mu}$ and ${z'}^{\mu}$,
\begin{equation}\label{eq:vp_final}
iV_0(x;x') = \frac{\mathcal{C}_0(D,a,b)}{(D\!-\!1) (D\!-\!2)} \times 
\frac{e^2 \kappa^2}{4} (\partial_{x} \downarrow - \partial_{x} \uparrow) 
\!\cdot\! (\partial_{x'} \downarrow - \partial_{x'} \uparrow) \times 
[i\Delta(x;x')]^2 \; .
\end{equation}
After applying the appropriate Donoghue Identity from Appendix C it turns out 
that all contributions to the amputated 4-scalar function take this same form, with 
different gauge dependent multiplicative factors. To simplify the notation, we 
define a new gauge dependent constant which includes the factor of $1/(D\!-\!1)
(D\!-\!2)$, and we take $D=4$ because dimensional regularization plays no role, 
while also dropping higher order perturbations in the gauge parameters $a = 1 + 
\delta a$ and $b = 1 + \delta b$,
\begin{equation}
\frac{\mathcal{C}_0(D,a,b)}{(D\!-\!1) (D\!-\!2)} \equiv C_0(\delta a,\delta b) +
O\Bigl(D\!-\!4,\delta^2\Bigr) \; . \label{newC0def}
\end{equation}
In other words, $C_0(\delta a,\delta b) = \frac43 + 2 \delta a$.

\subsection{Correlation between Vertices}
\begin{figure}[H]
\centering
\includegraphics[width=0.9\textwidth]{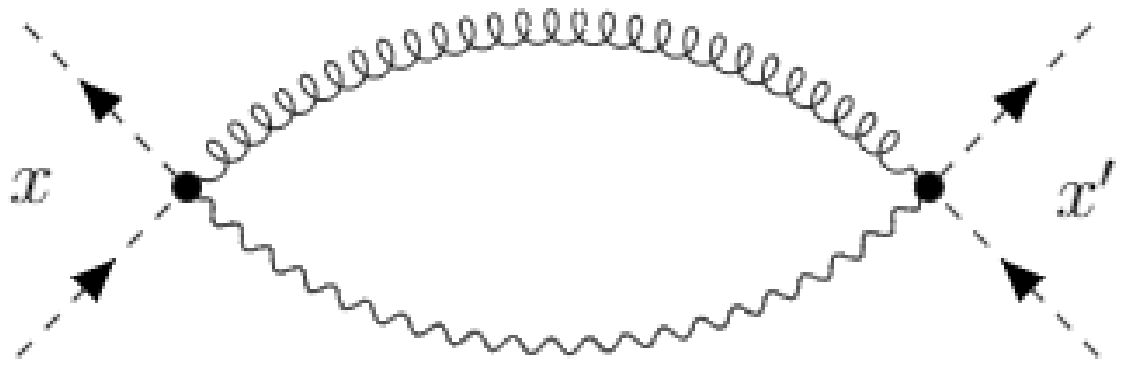}
\caption{\footnotesize This diagram shows the contribution of graviton correlation 
between two vertices. Dashed lines represent massive scalars, wavy lines represent 
photons and curly lines represent gravitons. These graphs have the same topology 
as Bjerrum-Bohr's Diagram 4 \cite{Bjerrum-Bohr:2002aqa}.}
\label{fig:feyn_betn_vert}
\end{figure}

The correlation between source (at ${x'}^{\mu}$) and observer (at $x^{\mu}$) 
vertices is the first extra contribution to the amputated 4-scalar function,
as shown in Figure \ref{fig:feyn_betn_vert}. This diagram corresponds to the
analytic expression,
\begin{equation}\label{eq:feyn_betn_vert}
\begin{split}
    iV_1(x;x')=&\frac{1}{2} e \kappa \left[\eta^{\delta\mu} \eta^{\nu\alpha}
    \!+\! \eta^{\delta\nu} \eta^{\mu\alpha} \!-\! \eta^{\mu\nu} \eta^{\delta\alpha}\right]
    (\partial_{x} \!\uparrow \!- \partial_{x}\!\downarrow)_{\alpha} \\
    & \times \frac{1}{2} e \kappa \left[\eta^{\gamma\rho} \eta^{\sigma\beta}
    \!+\! \eta^{\sigma\gamma} \eta^{\rho\beta} \!-\! \eta^{\rho\sigma} 
    \eta^{\gamma\beta}\right] (\partial_{x'} \!\uparrow \!- \partial_{x'}\!\downarrow)_{\beta} \\
    & \times i[_{\mu\nu}\Delta_{\rho\sigma}](x;x') \times i[_\gamma\Delta_\delta](x;x') \; .
\end{split}
\end{equation}
Substituting the appropriate propagators from Appendix B, contracting 
all the indices, simplifying and making use of the relation,
\begin{equation}\label{eq:prop_relation_1}
    i\Delta(x;x') \frac{\partial_\mu\partial_\nu}{\partial^2} i\Delta(x;x') = 
    \frac{1}{4} \!\times\! \frac{\Gamma^2(\frac{D}{2}-1)}{16\pi^D} 
    \left[\frac{\eta_{\mu\nu}}{\Delta x^{2D-4}} \!+\! \frac{\partial_{\mu}
    \partial_{\nu}}{(2D \!-\! 6)} \frac{1}{\Delta x^{2D-6}}\right] \; ,
\end{equation}
gives,
\begin{small}
\begin{equation}\label{eq:vert_corr_explicit}
    i V_1(x;x') = \frac{e^2 \kappa^2 \Gamma^2(\frac{D}{2} \!-\!1)}{16\pi^D}
    \left[D \!+\! \frac{(3D\!-\!2) \delta a}{4} \!-\! \frac{(D-2)^2 \delta b}{4}\right]
    (\partial_{x} \!\downarrow \!- \partial_{x} \!\uparrow)_{\mu} 
    (\partial_{x'} \!\downarrow \!- \partial_{x'} \!\uparrow)_{\nu} 
    \frac{\eta_{\mu\nu}}{\Delta x^{2D-4}} .
\end{equation}
\end{small}

\noindent Note that the second term in the square bracket of expression 
(\ref{eq:prop_relation_1}) drops out by current conservation.

Recognizing the massless scalar propagator (\ref{massless_scalar_prop}) provides 
a simpler form for (\ref{eq:vert_corr_explicit}),
\begin{equation}
i V_1(x;x') = \left[D \!+\! \frac{(3D\!-\!2) \delta a}{4} \!-\! \frac{(D-2)^2 
\delta b}{4}\right] \!\times\! e^2 \kappa^2 (\partial_{x} \!\downarrow \!- 
\partial_{x} \!\uparrow) \!\cdot\! (\partial_{x'} \!\downarrow \!- \partial_{x'}
\! \uparrow) \!\times\! [i\Delta(x;x')]^2 \; . \label{V1penult}
\end{equation}
As promised, expression (\ref{eq:vert_corr_explicit}) takes the same form as
the vacuum polarization contribution (\ref{eq:vp_final}), but with a different
gauge dependent, multiplicative constant. By comparison with (\ref{eq:vp_final})
we can recognize,
\begin{equation}
    \mathcal{C}_1(D,a,b) = 4 (D\!-\!1) (D\!-\!2)
    \left[D \!+\! \frac{(3D-2) \delta a}{4} \!-\! \frac{(D-2)^2 \delta b}{4}
    \!+\! O(\delta^2) \right] \; .
\end{equation}
Henceforth we will not bother with dimensional regularization, and we will make 
the same notational simplification as (\ref{newC0def}). This means that the
vertex-vertex correction is,
\begin{equation}\label{eq:vert_corr_final}
    i V_1(x;x') = C_1(\delta a,\delta b) \!\times\! \frac{e^2 \kappa^2}{4}
    (\partial_{x} \!\downarrow \!- \partial_{x}\! \uparrow) \!\cdot\! 
    (\partial_{x'} \!\downarrow \!- \partial_{x'}\! \uparrow)
    \!\times\! [i\Delta(x;x')]^2 \; ,
\end{equation}
where $C_1(\delta a,\delta b) = 16 + 10 \delta a - 4\delta b$.

\subsection{Vertex-Force Carrier Correlations}\label{sub_sec:vertex_force}

\begin{figure}[H]
\centering
\includegraphics[width=0.95\textwidth]{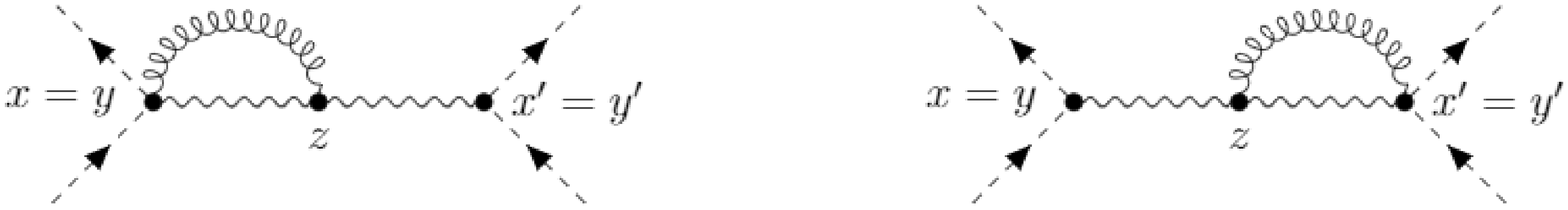}
\caption{\footnotesize These diagrams show the contributions from correlations between 
the force carrier and one of the vertices. Dashed lines represent massive scalars, wavy 
lines represent photons and curly lines represent gravitons. These graphs have the same 
topology as Bjerrum-Bohr's Diagram 7 \cite{Bjerrum-Bohr:2002aqa}.}
\label{fig:photon_vertices}
\end{figure}

The next contribution comes from the correlations between a single vertex and
the exchange photon, as shown in Figure \ref{fig:photon_vertices}. The 
analytic form is,
\begin{equation}
    \begin{split}
        iV_2(x;x')=&\frac{1}{2}e\kappa\left[\eta^{\epsilon\mu}\eta^{\nu\beta}+\eta^{\epsilon\nu}\eta^{\mu\beta}-\eta^{\mu\nu}\eta^{\epsilon\beta}\right](\partial_x\uparrow-\partial_x\downarrow)_\beta\times e(\partial_{x'}\downarrow-\partial_{x'}\uparrow)^\theta\\
        &\times\int d^Dz\ (-i\kappa V^{\gamma\delta\alpha\tau\rho\sigma})\ \partial_{z\tau}i[_\epsilon\Delta_\delta](x;z)\ \partial_{z\alpha}i[_\gamma\Delta_\theta](z;x')\times i[_{\mu\nu}\Delta_{\rho\sigma}](x;z)\\
        &+(\text{Permutation}) \; .
    \end{split}
\end{equation}
For reducing this diagram it is useful to note how the product of a
massless propagator times one of the gauge variations can be expressed
as a differential operator acting on a single function of the Poincar\'e
interval,
\begin{align}
    i\Delta(x;x')\frac{\partial_\mu\partial_\nu}{\partial^2}i\Delta(x;x')=&\frac{1}{4}\eta_{\mu\nu}[i\Delta(x;x')]^2-\frac{1}{8}\partial_\mu\partial_\nu I\{[i\Delta(x;x')]^2\} \; , \label{eq:id1} \\ 
    \begin{split}\label{eq:id2}
    \partial_\kappa i\Delta(x;x')\frac{\partial_\alpha\partial_\beta}{\partial^2}i\Delta(x;x')&=\frac{D-2}{16(D-1)}\partial_\kappa\partial_\alpha\partial_\beta I\{[i\Delta(x;x')]^2\}
        +\frac{D}{8(D-1)}\eta_{\alpha\beta}\partial_\kappa[i\Delta(x;x')]^2 \\
        &-\frac{D-2}{8(D-1)}(\eta_{\kappa\alpha}\partial_\beta
        + \eta_{\kappa\beta}\partial_\alpha)[i\Delta(x;x')]^2 \; ,
            \end{split}
\end{align}
where the symbol $I\{\}$ represents indefinite integration of the argument with 
respect to $\Delta x^2$. The final result for these diagrams is,
\begin{equation}
    i V_2(x;x') = C_2(\delta a,\delta b) \!\times\! \frac{e^2 \kappa^2}{4}
    (\partial_{x} \!\downarrow \!- \partial_{x}\! \uparrow) \!\cdot\! 
    (\partial_{x'} \!\downarrow \!- \partial_{x'}\! \uparrow)
    \!\times\! [i\Delta(x;x')]^2 \; ,
\end{equation}
where $C_2(\delta a,\delta b) = -12 -16 \delta a + 4 \delta b$.

\subsection{Vertex-Source and Vertex-Observer Correlations}\label{sec:triangular}

\begin{figure}[H]
\centering
\includegraphics[width=0.95\textwidth]{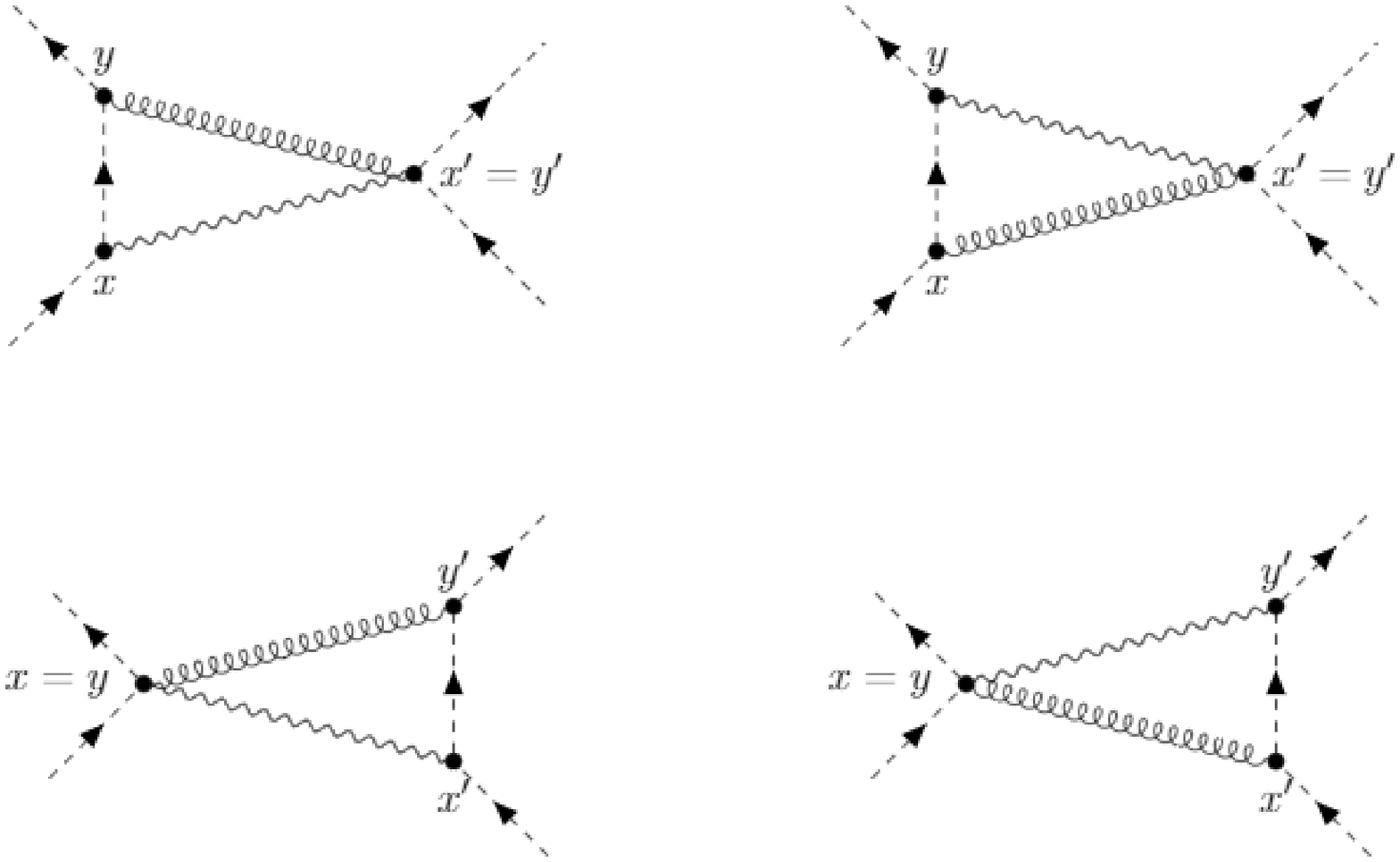}
\caption{\footnotesize These diagrams show the contributions from correlations 
between the source (primed) or observer (unprimed) and the opposite vertex. 
Dashed lines represent massive scalars, wavy lines represent photons and curly 
lines represent gravitons. These graphs have the same topology as Bjerrum-Bohr's 
Diagram 3 \cite{Bjerrum-Bohr:2002aqa}.}
\label{fig:triangular}
\end{figure}

We next consider contribution from correlations between the source, or observer,
and the opposite vertex, as shown in Figure \ref{fig:triangular}. (Correlations 
with nearer vertices do not contribute because they are cancelled by field 
strength renormalization.) We use $x^\mu$ ($y^\mu$) for incoming (outgoing) 
observer, and ${x'}^\mu$ (${y'}^\mu$) for incoming (outgoing) source. We also
adopt the notation that a bar over a vertex with only a single external leg
denotes differentiation of the on-shell external wave function. With these
conventions we can write the analytic form of the diagrams in 
Figure~\ref{fig:triangular} as,
\begin{equation} \label{triangular}
    \begin{split}
        {\rm Figure\ \ref{fig:triangular}} =& \frac{i\kappa}{2} \left[
        \bar{\partial}^{\mu}_y \partial^{\nu}_y
        \!+\! \bar{\partial}^{\nu}_y \partial^{\mu}_y \!-\! \eta^{\mu\nu} \left( 
        \partial_{y} \!\cdot\! \bar{\partial}_{y} \!+\! m^2\right)\right] \!\times\!
        e(\bar{\partial}_{x} \!-\! \partial_{x})^{\delta} i\Delta_m(x;y) \\
        & \times \frac{e\kappa}{2} \left[\eta^{\gamma\rho} \eta^{\alpha\sigma} \!+\!
        \eta^{\gamma\sigma} \eta^{\alpha\rho} \!-\! \eta^{\rho\sigma} \eta^{\alpha\gamma}
        \right](\partial_{x'} \!\uparrow \!- \partial_{x'}\!\downarrow)_{\alpha} \!\times\! 
        i[_{\delta} \Delta_{\gamma}](x;x') \!\times\! i[_{\mu\nu} \Delta_{\rho\sigma}](y;x') \\
        & + (3\ \text{permutations}) \; .
    \end{split}
\end{equation}

As can be seen from Figure~\ref{fig:triangular}, these contributions involve an 
internal massive scalar propagator in the loop. This poses an obstacle to regarding 
expression (\ref{triangular}) as a correction to the vacuum polarization. This is 
overcome through the ``Donoghue Identities'' of Appendix \ref{sec:donoghue},
which degenerate the massive scalar propagator to a Dirac delta function, and 
reduce expression (\ref{triangular}) to the same 2-point form (\ref{eq:vp_final})
as the contribution from the original vacuum polarization. The part of expression
(\ref{triangular}) which is independent of the gauge parameters $\delta a$ and
$\delta b$ reaches the desired form through the Donoghue Identities (\ref{eq:3pt})
and (\ref{eq:3pt_derivative}). 

As an example of the part of (\ref{triangular}) proportional to $\delta a$, we 
consider the term,
\begin{equation}\label{eq:trick1}
    \begin{split}
        \delta a & \times \frac{i\kappa}{2} \left[\bar{\partial}^{\mu}_y
        \partial^{\nu}_y \!+\! \bar{\partial}^{\nu}_y \partial^{\mu}_y \!-\!
        \eta^{\mu\nu} \left(\partial_y \!\cdot\! \bar{\partial}_y \!+\! m^2\right)
        \right] \!\times\! e(\bar{\partial}_x \!-\! \partial_x)^{\delta} i\Delta_m(x;y) \\
        & \times \frac{e\kappa}{2} \left[\eta^{\gamma\rho} \eta^{\alpha\sigma}
        \!+\! \eta^{\gamma\sigma} \eta^{\alpha\rho} \!-\! \eta^{\rho\sigma}
        \eta^{\alpha\gamma}\right](\partial_{x'} \!\uparrow \!- \partial_{x'}\! 
        \downarrow)_{\alpha} \!\times\! i[_{\delta} \Delta_{\gamma}](x;x') \!\times\!
        \eta_{\nu\rho} \frac{\partial_{\mu} \partial_{\sigma}} {\partial^2} 
        i \Delta(y;x') \; .
    \end{split}
\end{equation}
The derivative $\partial_\mu$ acting on the massless propagator $i\Delta(y;x')$ can 
be partially integrated to act on the massive propagator $i\Delta_m(x;y)$,
\begin{equation}\label{eq:conservation_trick}
    -(\bar{\partial}_y \!+\! \partial_y)_{\mu} \left[ \bar{\partial}^{\mu}_y
    \partial^{\nu}_y \!+\! \bar{\partial}^{\nu}_y \partial^{\mu}_y \!-\!
    \eta^{\mu\nu}\left(\partial_y \!\cdot\! \bar{\partial}_y \!+\! m^2\right)
    \right] = -\bar{\partial}_y^{\nu} i\delta^D(x \!-\! y) \; .
\end{equation}
The remaining factor of $\partial^{\sigma}/\partial^2$ can be reduced using,
\begin{align}
    & i\Delta(x;x') \!\times\! \frac{\partial_\alpha}{\partial^2} i\Delta(x;x')
    = \frac{1}{4} \partial_{\alpha} I\{[i\Delta(x;x')]^2\} \; , \\
    & \partial^2 I\{[i\Delta(x;x')]^2\} \longrightarrow -2 (D\!-\!4) 
    [i\Delta(x;x')]^2 \; .
\end{align}

From expression (\ref{eq:graviton_prop}) for the graviton propagator we see that
there are two parts proportional to $\delta b$. The second term with $\partial_{\mu}
\partial_{\nu}/\partial^2$ can be reduced just like (\ref{eq:trick1}). The other 
term requires additional effort,
\begin{equation}\label{eq:trick2}
    \begin{split}
        -2 \delta b & \times \frac{i\kappa}{2} \left[\bar{\partial}^{\mu}_y 
        \partial^{\nu}_y \!+\! \bar{\partial}^{\nu}_y \partial^{\mu}_y \!-\!
        \eta^{\mu\nu} \left(\partial_y \!\cdot\! \bar{\partial}_y \!+\! m^2\right)
        \right] \!\times\! e(\bar{\partial}_x \!-\! \partial_x)^{\delta} 
        i\Delta_m(x;y) \\
        & \times \frac{e\kappa}{2} \left[\eta^{\gamma\rho} \eta^{\alpha\sigma} \!+\!
        \eta^{\gamma\sigma} \eta^{\alpha\rho} \!-\! \eta^{\rho\sigma} \eta^{\alpha\gamma}
        \right](\partial_{x'}\! \uparrow \!- \partial_{x'} \!\downarrow)_{\alpha}
        \!\times\! i[_{\delta}\Delta_{\gamma}](x;x') \!\times\! \eta_{\mu\nu}
        \frac{\partial_{\rho} \partial_{\sigma}}{\partial^2} i \Delta(y;x') \; .
    \end{split}
\end{equation}
To reduce this term we distinguish between $y^{\mu}$ derivatives acting on the
external leg ($\bar{\partial}_{y}^{\mu}$), the massive propagator ($\partial_{y}^{\mu}$)
and the massless propagator of the graviton ($\tilde{\partial}_{y}^{\mu}$),
\begin{equation}
\bar{\partial}_{y}^{\mu} + \partial_{y}^{\mu} + \tilde{\partial}_{y}^{\mu} = 0 \; .
\end{equation}
Now note that,
\begin{equation}\label{eq:trick3}
    \eta_{\mu\nu} \left[\bar{\partial}^{\mu}_y \partial^{\nu}_y \!+\! 
    \bar{\partial}^{\nu}_y \partial^{\mu}_y \!-\! \eta^{\mu\nu} \left(\partial_y
    \!\cdot\! \bar{\partial}_y \!+\! m^2\right) \right] = (\bar{\partial}_y^2 \!-\! m^2)
    + (\partial_y^2 \!-\! m^2) - \tilde{\partial}_y^2 - 2 m^2 \; .
\end{equation}
The factor of $(\bar{\partial}_y^2 - m^2)$ vanishes due to the external leg being on shell.
The next term in (\ref{eq:trick3}) degenerates the massive scalar propagator, 
\begin{equation}
    (\partial_y^2 \!-\! m^2) i\Delta_m(x;y) = i \delta^D(x \!-\! y) \; .
\end{equation} 
Of course the factor of $\tilde{\partial}_y^2$ eliminates the troublesome inverse 
D`Alembertian, whereupon the Donoghue Identity (\ref{eq:3pt_2derivative}) completes the
reduction. The final term in (\ref{eq:trick3}) requires the newly Donoghue Identities 
(\ref{eq:new_dono_1}) and (\ref{eq:new_dono_2}). 

Putting everything together gives the final result for Figure~\ref{fig:triangular},
\begin{equation}
    i V_3(x;x') = C_3(\delta a,\delta b) \!\times\! \frac{e^2 \kappa^2}{4}
    (\partial_{x} \!\downarrow \!- \partial_{x}\! \uparrow) \!\cdot\! 
    (\partial_{x'} \!\downarrow \!- \partial_{x'}\! \uparrow)
    \!\times\! [i\Delta(x;x')]^2 \; ,
\end{equation}
where $C_3(\delta a,\delta b) = -32 -8 \delta a - 2 \delta b$.

\subsection{Source-Observer Correlations}
\begin{figure}[H]
\centering
\includegraphics[width=1\textwidth]{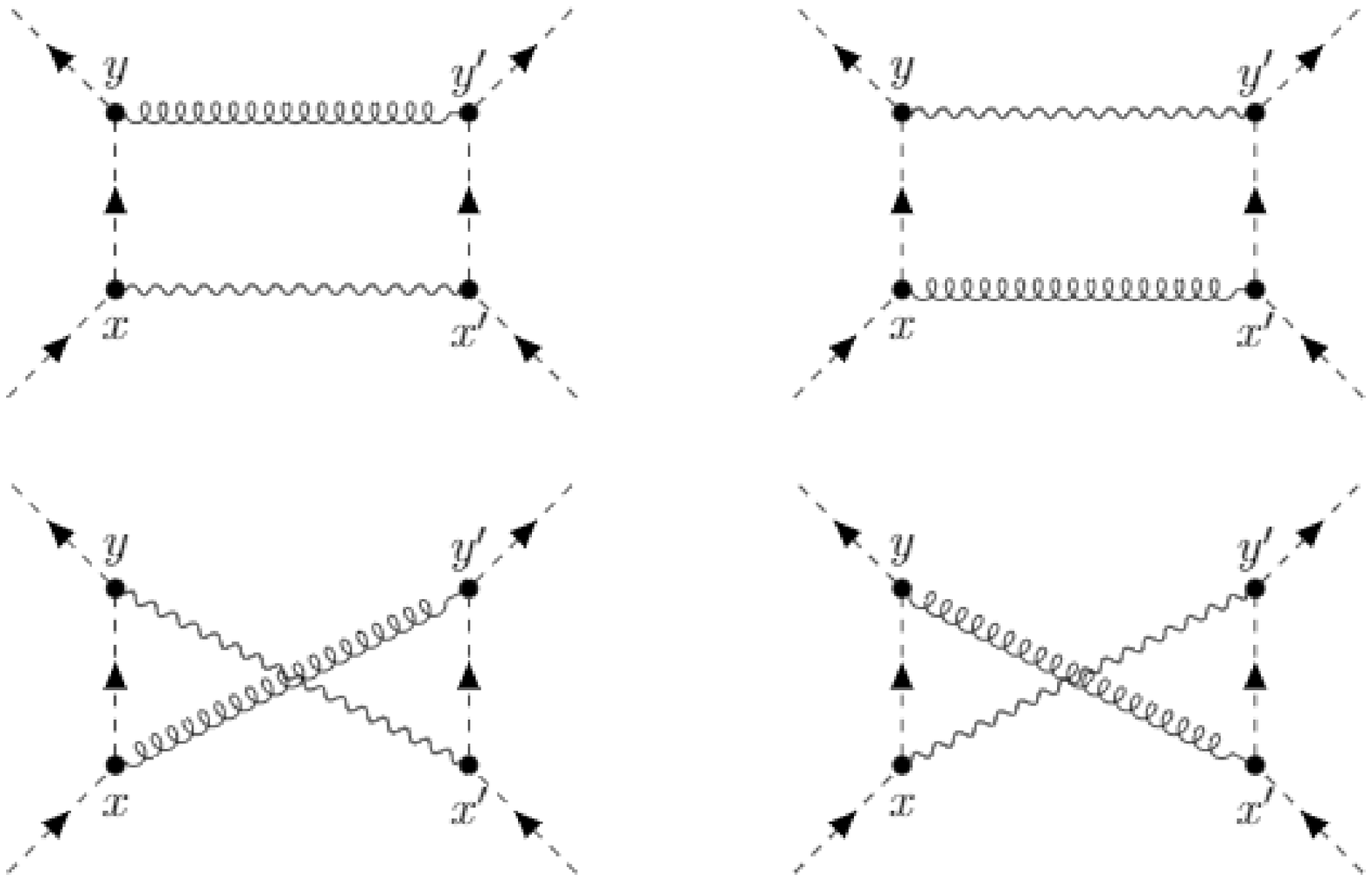}
\caption{\footnotesize These diagrams show the contributions from correlations 
between the source (primed) and observer (unprimed). Dashed lines represent 
massive scalars, wavy lines represent photons and curly lines represent 
gravitons. These graphs have the same topology as Bjerrum-Bohr's Diagram 2
\cite{Bjerrum-Bohr:2002aqa}.}
\label{fig:box}
\end{figure}

Figure~\ref{fig:box} shows contributions from correlations between source and 
observer. (Correlations between the source and itself, or the observer and itself,
do not correct the exchange photon.) The analytic form for these diagrams is,
\begin{equation} \label{box}
    \begin{split}
        (\rm Figure\ \ref{fig:box}) = & \frac{i\kappa}{2} \left[\bar{\partial}_y^{\mu}
        \partial_y^{\nu} \!+\! \bar{\partial}_y^{\nu} \partial_y^{\mu} \!-\!
        \eta^{\mu\nu} \left(\bar{\partial}_y \!\cdot\! \partial_y \!+\! m^2\right) 
        \right]\left(\bar{\partial}_x \!-\! \partial_x\right)^{\delta} i\Delta_m(x;y) \\
        & \times \frac{i\kappa}{2} \left[\bar{\partial}_{y'}^{\rho} \partial_{y'}^{\sigma}
        \!+\! \bar{\partial}_{y'}^{\sigma} \partial_{y'}^{\rho} \!-\! \eta^{\rho\sigma}
        \left(\bar{\partial}_{y'} \!\cdot\! \partial_{y'} \!+\! m^2\right) \right]
        \left(\bar{\partial}_{x'} \!-\! \partial_{x'}\right)^{\gamma} i\Delta_m(x';y') \\
        & \times i[_{\gamma} \Delta_{\delta}](x;x') \!\times\! i[_{\mu\nu} 
        \Delta_{\rho\sigma}](y;y') + (3 \text{ Permutations}) \; .
    \end{split}
\end{equation}
Note that the two permutations on the bottom line of Figure~\ref{fig:box} contain an
extra minus sign due to 2-scalar-1-photon vertex. 

The part of (\ref{box}) independent of $\delta a$ and $\delta b$ is accomplished by the
Donoghue Identity (\ref{eq:box_dono}). The reduction of the gauge dependent parts is
similar to what we have seen before with one difference: after using relation 
(\ref{eq:conservation_trick}), one must combine parts from the various diagrams to 
eliminate some troublesome terms. The final result for Figure~\ref{fig:box} is,
\begin{equation}
    i V_4(x;x') = C_4(\delta a,\delta b) \!\times\! \frac{e^2 \kappa^2}{4}
    (\partial_{x} \!\downarrow \!- \partial_{x}\! \uparrow) \!\cdot\! 
    (\partial_{x'} \!\downarrow \!- \partial_{x'}\! \uparrow)
    \!\times\! [i\Delta(x;x')]^2 \; ,
\end{equation}
where $C_4(\delta a,\delta b) = \frac{80}{3} + 0 \cdot \delta a +2 \delta b$.

\subsection{Force Carrier Correlations with Source and Observer}
\begin{figure}[H]
\centering
\includegraphics[width=0.95\textwidth]{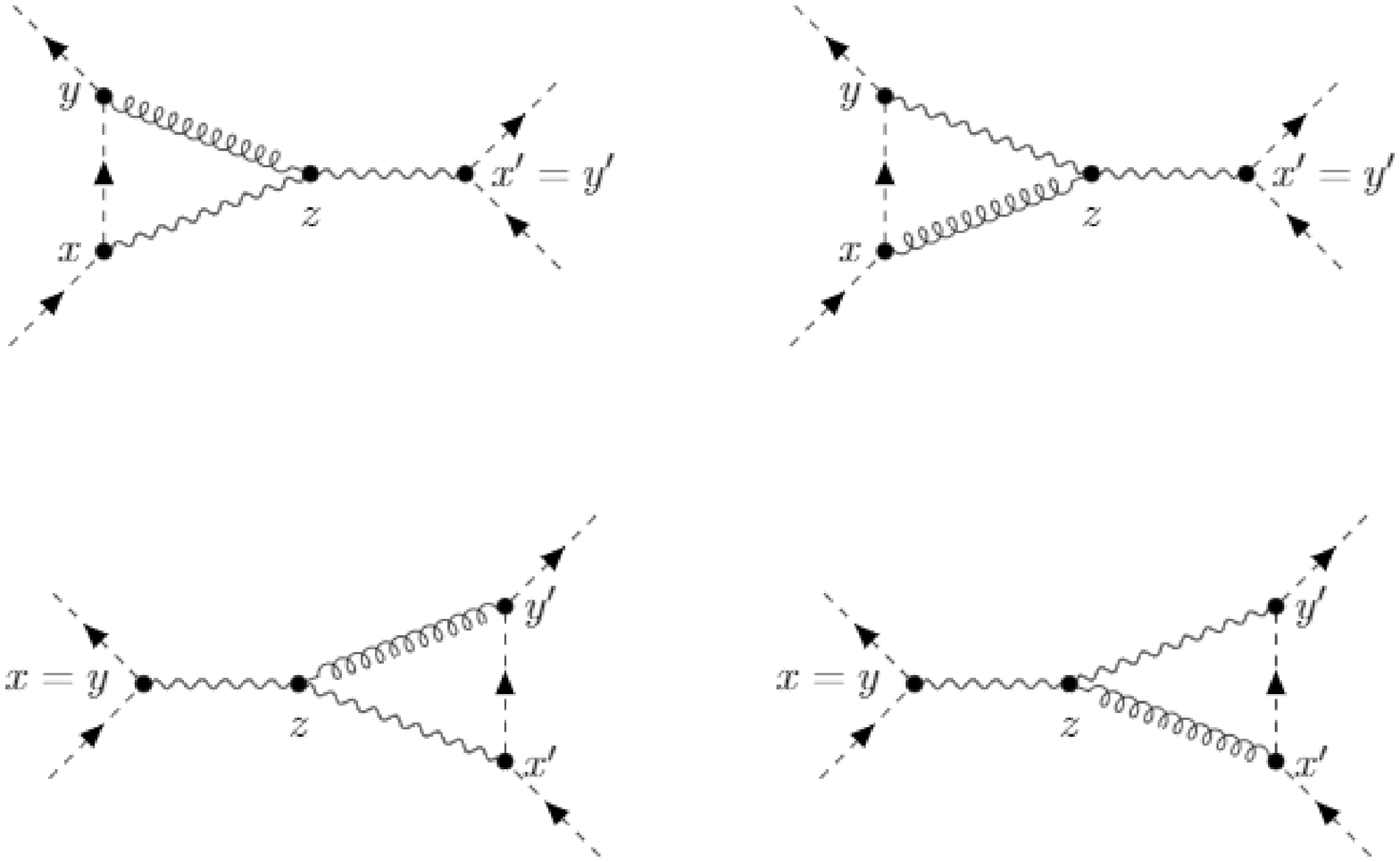}
\caption{\footnotesize These diagrams show the contributions from correlations 
between the source (primed) or observer (unprimed) and the force carrier. Dashed 
lines represent massive scalars, wavy lines represent photons and curly lines 
represent gravitons. These graphs have the same topology as Bjerrum-Bohr's 
Diagram 6 \cite{Bjerrum-Bohr:2002aqa}.}
\label{fig:force_source_obs}
\end{figure}

The next contribution comes from correlations between the source or observer and
the photon. The Feynman diagrams are given in Figure~\ref{fig:force_source_obs},
and the analytic expression is,
\begin{equation}
    \begin{split}
        {\rm (Figure\ \ref{fig:force_source_obs})} = & \frac{i\kappa}{2} \left[
        \bar{\partial}_y^{\mu} \partial_y^{\nu} \!+\! \bar{\partial}_y^{\nu}
        \partial_y^{\mu} \!-\! \eta^{\mu\nu} \left(\bar{\partial}_y \!\cdot\!
        \partial_y \!+\! m^2\right) \right] \left(\bar{\partial}_x \!-\! 
        \partial_x\right)^{\epsilon} i\Delta_m(x;y) \\
        & \times \int \!\! d^Dz (-i\kappa V^{\gamma\delta\alpha\tau\rho\sigma}) 
        \frac{\partial}{\partial z^{\tau}} i[_{\epsilon} \Delta_{\delta}](x;z) 
        \frac{\partial}{\partial z^{\alpha}} i[_{\gamma} \Delta_{\theta}](z;x') \!\times\!
        e\left(\partial_{x'} \!\downarrow \!- \partial_{x'} \!\uparrow\right)^\theta \\
        & \times i[_{\mu\nu} \Delta_{\rho\sigma}] + (3\text{ Permutations}) \; .
    \end{split}
\end{equation}
The reduction process is almost same as in Section \ref{sec:triangular}, the chief
difference being the extra photon propagator. We extract a D'Alembertian and then use 
(\ref{eq:equation_massless}) to eliminate this and the integration over $z^{\mu}$. 
The final contribution from these diagrams is,
\begin{equation}
    i V_5(x;x') = C_5(\delta a,\delta b) \!\times\! \frac{e^2 \kappa^2}{4}
    (\partial_{x} \!\downarrow \!- \partial_{x}\! \uparrow) \!\cdot\! 
    (\partial_{x'} \!\downarrow \!- \partial_{x'}\! \uparrow)
    \!\times\! [i\Delta(x;x')]^2 \; ,
\end{equation}
where $C_5(\delta a,\delta b) = 12 + 12 \delta a + 4 \delta b$.

\subsection{Gravitational 1-PR Vertex Corrections}

\begin{figure}[H]
\centering
\includegraphics[width=0.95\textwidth]{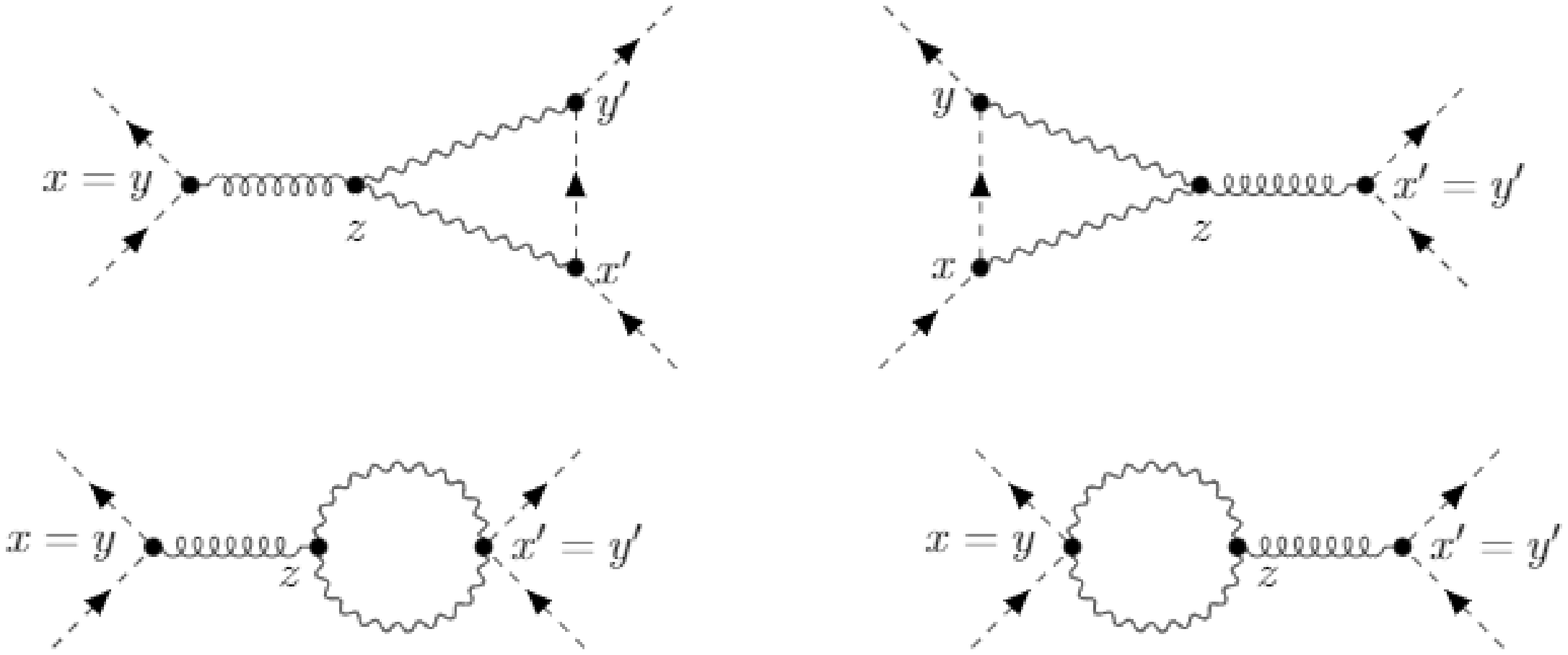}
\caption{\footnotesize These diagrams show the contributions from the 1PR
(one particle reducible) diagrams corresponding to gravitational vertex 
corrections. Dashed lines represent massive scalars, wavy lines represent 
photons and curly lines represent gravitons. These graphs have the same 
topology as Bjerrum-Bohr's Diagram 5 \cite{Bjerrum-Bohr:2002aqa}.}
\label{fig:graviton_vertex}
\end{figure}

The final contribution to the amputated 4-scalar function comes from 
diagrams in which a loop of photons corrects one of the vertices and
the graviton carries the exchange force. The relevant diagrams are shown 
in Figure~\ref{fig:graviton_vertex}.
The analytic form for first two diagrams is,
\begin{equation}
    \begin{split}
        \text{Figure 6}_\text{up} = & \frac{i\kappa}{2} \left[\partial_x^{\mu} \!\uparrow
        \partial_x^{\nu} \! \downarrow \!+\! \partial_x^{\nu} \!\uparrow
        \partial_x^{\mu} \!\downarrow \!-\! \eta^{\mu\nu} \left(\partial_x
        \!\uparrow \!\cdot\! \partial_x \!\downarrow \!+\! m^2\right)\right]
        \!\times\! e\left(\bar{\partial}_{x'} \!-\! \partial_{x'}\right)^{\theta} \\
        & \times e\left(\partial_{y'} \!-\! \bar{\partial}_{y'}\right)^{\epsilon} 
        i\Delta_m(x';y') \int\!\! d^Dz\ (-i\kappa V^{\gamma\delta\alpha\tau\rho\sigma})
        \frac{\partial}{\partial z^{\tau}} i[_{\epsilon}\Delta_{\delta}](z;x') 
        \frac{\partial}{\partial z^{\alpha}} i[_\gamma\Delta_\theta](z;x') \\
        & \times i[_{\mu\nu} \Delta_{\rho\sigma}](x;z) + (\text{Permutation}) \; .
    \end{split}
\end{equation}
The second two diagrams are,
\begin{equation}
    \begin{split}
        \text{Figure 6}_\text{down} = & \frac{1}{2} \!\times\! \frac{i\kappa}{2} \left[\partial_x^{\mu}
        \!\uparrow \partial_x^{\nu} \!\downarrow \!+\! \partial_x^{\nu} \!\uparrow
        \partial_x^{\mu} \! \downarrow \!-\! \eta^{\mu\nu} \left(\partial_x \!\uparrow
        \!\cdot\! \partial_x \!\downarrow \!+\! m^2\right)\right] \!\times\! (-2 i e^2
        \eta^{\epsilon\theta}) \\
        &\int \!\! d^Dz (-i\kappa V^{\gamma\delta\alpha\tau\rho\sigma}) 
        \frac{\partial}{\partial z^{\tau}} i[_{\epsilon}\Delta_{\delta}](z;y')
        \frac{\partial}{\partial z^{\alpha}} i[_{\gamma} \Delta_{\theta}](z;x') \\
        & \times i[_{\mu\nu}\Delta_{\rho\sigma}](x;z) + (\text{Permutation}) \; .
    \end{split}
\end{equation}
Whenever possible, it is prudent to partially integrate and reflect a derivative 
through the graviton propagator to act on the 2-scalar-1-graviton vertex. One then makes
use of the relation,
\begin{equation}
    (\partial_x \!\downarrow \!+\! \partial_x\! \uparrow)_{\mu} \left[\partial_x^{\mu}
    \! \uparrow \partial_x^{\nu} \!\downarrow \!+\! \partial_x^{\nu} \!\uparrow
    \partial_x^{\mu} \!\downarrow \!-\! \eta^{\mu\nu} \left(\partial_x \!\uparrow \!\cdot\!
    \partial_x \!\downarrow \!+\! m^2\right)\right] = 0 \; .
\end{equation}
The rest of the reduction is same as in earlier sections. The final result is,
\begin{equation}
    i V_6(x;x') = C_6(\delta a,\delta b) \!\times\! \frac{e^2 \kappa^2}{4}
    (\partial_{x} \!\downarrow \!- \partial_{x}\! \uparrow) \!\cdot\! 
    (\partial_{x'} \!\downarrow \!- \partial_{x'}\! \uparrow)
    \!\times\! [i\Delta(x;x')]^2 \; ,
\end{equation}
where $C_6(\delta a,\delta b) = -\frac{16}{3} + 0 \cdot \delta a - 4 \delta b$.

\subsection{Sum Total}

As we have explained, the Donoghue Identities of Appendix C allow us to cast each 
contribution to the amputated 4-scalar function in the form,
\begin{equation}
    i V_i(x;x') = C_i(\delta a,\delta b) \!\times\! \frac{e^2 \kappa^2}{4} 
    (\partial_{x} \!\downarrow \!- \partial_{x}\! \uparrow) \!\cdot\! 
    (\partial_{x'} \!\downarrow \!- \partial_{x'}\! \uparrow)
    \!\times\! [i\Delta(x;x')]^2 \; ,
\end{equation}
where the gauge dependent constant is $C_i(\delta a,\delta b) =O_i + A_i \delta a
+ B_i \delta b$. Table~\ref{tab:sum_total} summarizes our results.
\begin{table}[H]
    \centering
\begin{tabular}[c]{||c|c|c|c|c||}
\hline
\rule{0pt}{18pt} $i$ & Description & $\quad O_i\quad $& $\quad A_i\quad$&$\quad B_i\quad$\\[8pt]
\hline\hline
     \rule{0pt}{18pt}0 & Vacuum Polarization & $+\frac{4}{3}$ & $+2$ & $0$\\[8pt]
     \hline
     \rule{0pt}{18pt}1 & Circular Diagram & $+16$ & $+10$ & $-4$\\[8pt]
     \hline
     \rule{0pt}{18pt}2 & Vertex-Force Carrier & $-12$ & $-16$ & $+4$\\[8pt]
     \hline
     \rule{0pt}{18pt}3 & Triangular Diagrams & $-32$ & $-8$ & $-2$\\[8pt]
     \hline
     \rule{0pt}{18pt}4 & Box Diagrams & $+\frac{80}{3}$ & $0$ & $+2$\\[8pt]
     \hline
     \rule{0pt}{18pt}5 & Source, Obs.- Force Carrier & $+12$ & $+12$ & $+4$\\[8pt]
     \hline
     \rule{0pt}{18pt}6 & Graviton 1PR Vertex & $-\frac{16}{3}$ & $0$ & $-4$\\[8pt]
     \hline\hline
     \rule{0pt}{18pt} & Total & $+\frac{20}{3}$ & $0$ & $0$\\[8pt]
      \hline
\end{tabular}
\caption{ The entry on the $i^{\text{th}}$ row represent the gauge dependent factors for 
the contribution coming from the diagram in Figure $i$.}
    \label{tab:sum_total}
\end{table}
Of course the point of the exercise is to total the $O_i$'s, and to show that the
terms proportional to $\delta a$ and $\delta b$ sum to zero,
\begin{equation}
    \sum_{i=0}^6 i V_{i}(x;x') = +\frac{20}{3} \times \frac{e^2 \kappa^2}{4}
    (\partial_{x} \!\downarrow \!- \partial_{x}\! \uparrow) \!\cdot\! 
    (\partial_{x'} \!\downarrow \!- \partial_{x'}\! \uparrow)
    \!\times\! [i\Delta(x;x')]^2 \; .
\end{equation}
We then reverse the steps that led from expression (\ref{eq:so}) to (\ref{eq:vp_final})
in order to conclude that the gauge independent vacuum polarization from a single
graviton loop is,
\begin{equation}
    i[^{\mu} \Pi^{\nu}](x;x') = -\frac{40}{3} \!\times\!
    \frac{\kappa^2}{16\pi^4} \left[\eta^{\mu\nu} \partial^2 \!-\! \partial^{\mu}
    \partial^{\nu}\right] \frac{1}{\Delta x^{2D-2}} \; .
\end{equation}

\section{Conclusions}

The main result of this paper is that including quantum gravitational
corrections from the source which disturbs the effective field, and from
the observer who measures the disturbance, eliminates the massive gauge 
dependence of the quantum-corrected Maxwell equation (\ref{QMax}) that
was evident in the multiplicative constant $\mathcal{C}_0(D,a,b)$ of 
expression (\ref{C0def}). After renormalization, and application of the 
Schwinger-Keldysh formalism \cite{Ford:2004wc}, our final result for the 
one loop effective field equation is,
\begin{equation}
\partial_{\nu} F^{\nu\mu}(x) + \frac{5 \hbar G \partial^6}{48 \pi^2 c^3} 
\! \int \!\! d^4x' \, \theta(\Delta t \!-\! \Delta r) \Bigl\{ \ln[\mu^2 
(\Delta t^2 \!-\! r^2)] \!-\! 1\Bigr\} \partial_{\nu}' F^{\nu\mu}(x') = 
J^{\mu}(x) \; , \label{QMaxfinal}
\end{equation}
where $\Delta t \equiv t - t'$, $r \equiv \Vert \vec{x} - \vec{x}'\Vert$
and we have restored the factors $\hbar$ and $c$. Although equation 
(\ref{QMaxfinal}) is not local, it is real and causal.

For a static point charge $J^{\mu}(t,\vec{x}) = q \delta^{\mu}_{~0} 
\delta^3(\vec{x} \!-\! \vec{x}')$ the quantum-corrected Coulomb potential is,
\begin{equation}
\Phi(r) = \frac{q}{4\pi r} \left\{ 1 + \frac{10 \hbar G}{3\pi r^2 c^3}
+ \mathcal{O}(G^2) \right\} .
\end{equation}
This result agreees with Bjerrum-Bohr \cite{Bjerrum-Bohr:2002aqa}, but we now
have the ability to solve for quantum gravitational corrections to the full
range of problems one encounters in classical electrodynamics. These 
corrections are bound to be quite small under ordinary conditions, although 
the potential for slightly super-luminal propagation is noteworthy 
\cite{Leonard:2012fs}, and was predicted long ago \cite{Deser:1957zz,
DeWitt:1960fc}.

Although it is nice to finally be able to include quantum gravitational 
corrections to Maxwell's equations on flat space background, we could always 
have inferred physics from scattering amplitudes. The real necessity for our 
method is for studying quantum gravitational corrections to electrodynamics 
in cosmology. These effects can be significant, especially during the epoch 
of primordial inflation. For example, when the simplest de Sitter background
gauge \cite{Tsamis:1992xa,Woodard:2004ut} is employed to compute single
graviton loop corrections to the vacuum polarization \cite{Leonard:2013xsa}
one finds corrections to the Coulomb potential \cite{Glavan:2013jca}, and
to the photon field strength \cite{Wang:2014tza} which become nonperturbatively
strong at large distances and late times. When the vacuum polarization is 
computed in a much more complicated, 1-parameter family of gauges 
\cite{Glavan:2015ura}, one finds the same time dependence for the photon 
field strength, but with a different numerical coefficient \cite{Glavan:2016bvp},
signaling a slight gauge dependence which must be eliminated to infer reliable
results. First order corrections to the graviton propagator in the de Sitter 
generalization of the gauge (\ref{gauge}) have been derived recently 
\cite{Glavan:2019msf}. This should facilitate extending the current work
to de Sitter background.

\vskip 1cm

\centerline{\bf Acknowledgements}

We are grateful to J. F. Donoghue for correspondence on this 
subject. This work was partially supported by by NSF grant 
PHY-1912484 and by the Institute for Fundamental Theory at 
the University of Florida.

\appendix
\numberwithin{equation}{section}
\section{Appendix: The Vertices}

\begin{tabular}{ll}
\textit{$\bullet$ 2-Scalars-1-Photon Vertex} & \\
\parbox[c]{4cm}{
      \includegraphics[width=4cm]{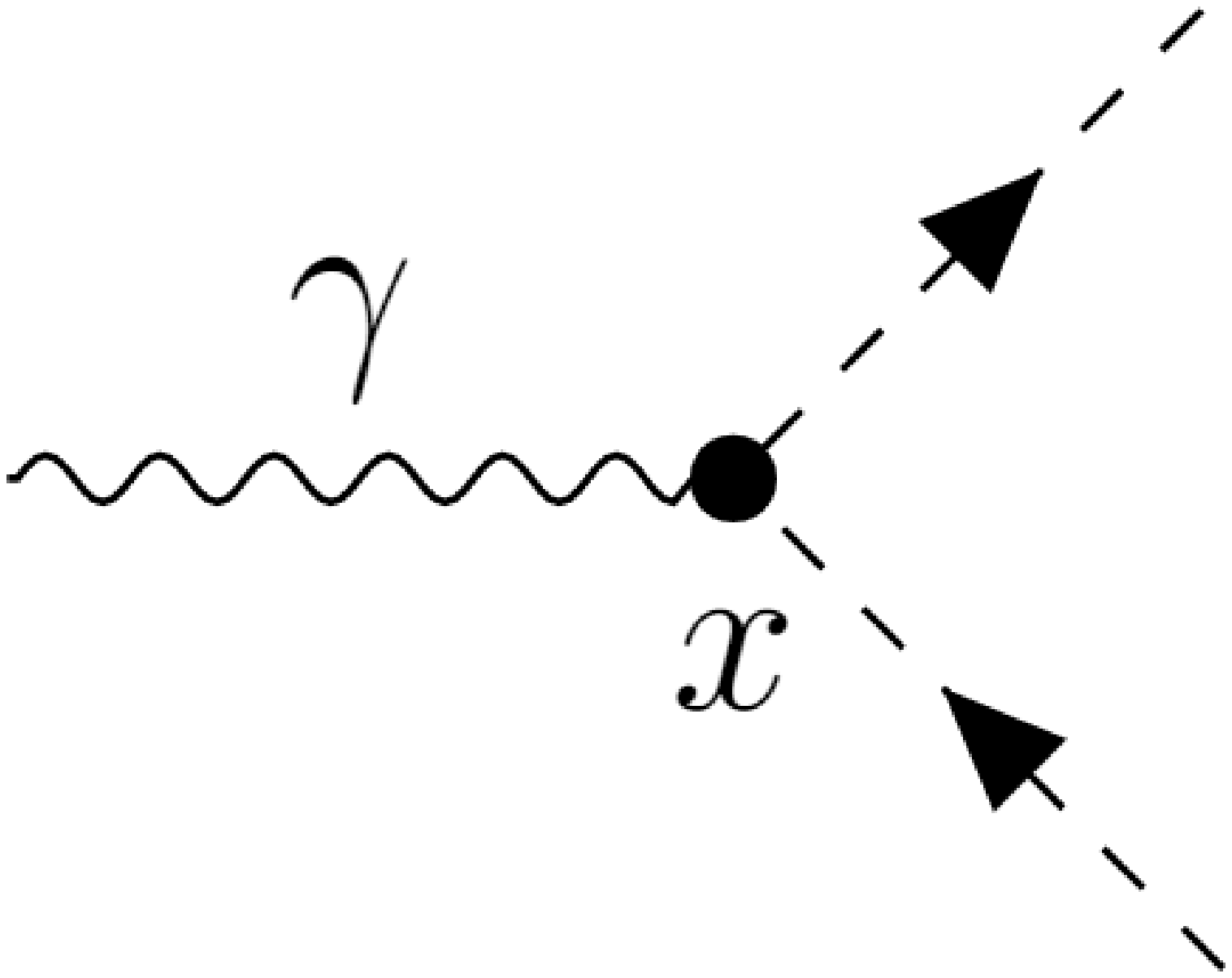}} & \parbox[l]{8.4cm}{\begin{equation} =e(\partial_x\downarrow-\partial_x\uparrow)^\gamma 
      \end{equation}}\\[2cm]
   \textit{$\bullet$ 2-Scalars-2-Photon Vertex}& \\
   \parbox[c]{4cm}{\includegraphics[width=4cm]{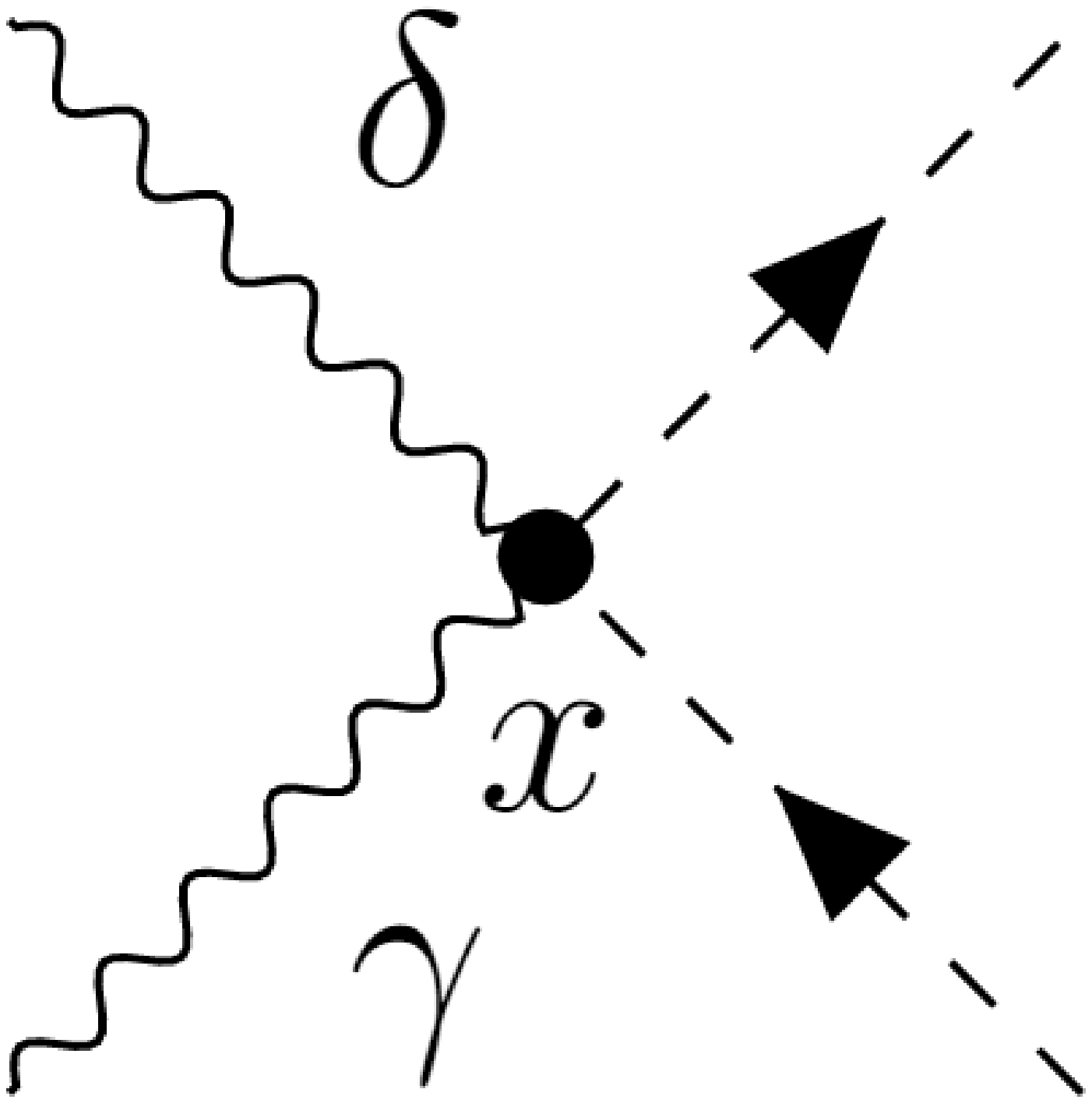}} & \parbox[l]{8.4cm}{\begin{equation}=-2ie^2\eta^{\gamma\delta}
   \end{equation}}\\[2cm]
   \textit{$\bullet$ 2-Scalars-1-Graviton Vertex} & \\
   \parbox[c]{4cm}{\includegraphics[width=4cm]{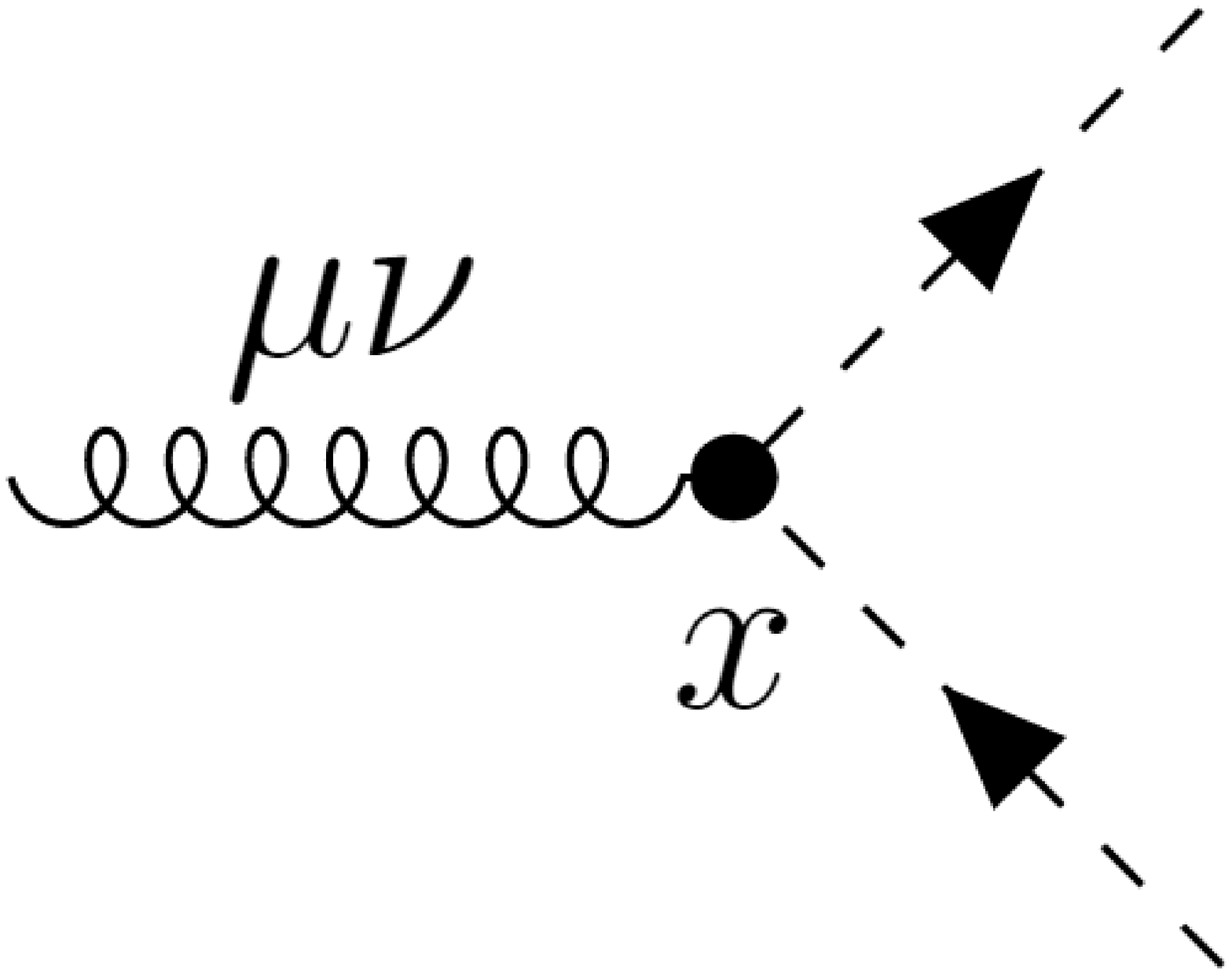}} &  \parbox[l]{8.4cm}{\begin{equation}\begin{split}
=\frac{i\kappa}{2}[&\partial_x^\mu\uparrow\partial^\nu_x\downarrow+\partial^\nu_x\uparrow\partial^\mu_x\downarrow\\
&- \eta^{\mu\nu}(\partial_x\uparrow\cdot\partial_x\downarrow+m^2)]
\end{split}
 \end{equation}}\\[2cm]
   \textit{$\bullet$ 2-Scalars-1-Photon-1-Graviton Vertex} & \\
   \parbox[c]{4cm}{\includegraphics[width=4cm]{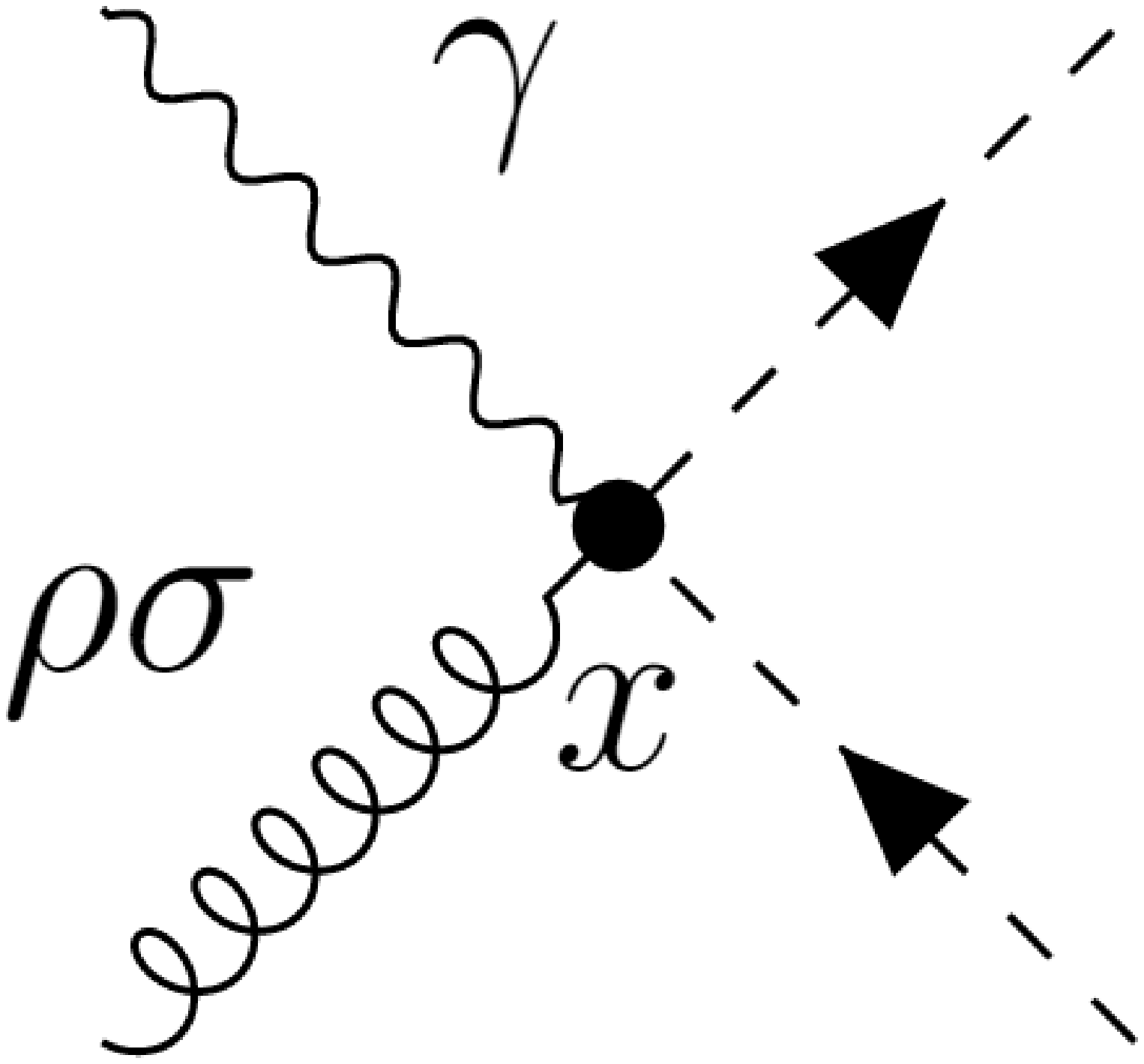}} & \parbox[l]{8.4cm}{\begin{equation}\begin{split}
   =\frac{e\kappa}{2}[&\eta^{\gamma\rho}\eta^{\alpha\sigma}+\eta^{\gamma\sigma}\eta^{\alpha\rho}-\eta^{\rho\sigma}\eta^{\alpha\gamma}]\\
   &\times(\partial_x\uparrow-\partial_x\downarrow)_\alpha
   \end{split}
   \end{equation}}\\[2cm]
   \textit{$\bullet$ 2-Photon-1-Graviton Vertex} & \\
   \parbox[c]{4cm}{\includegraphics[width=4cm]{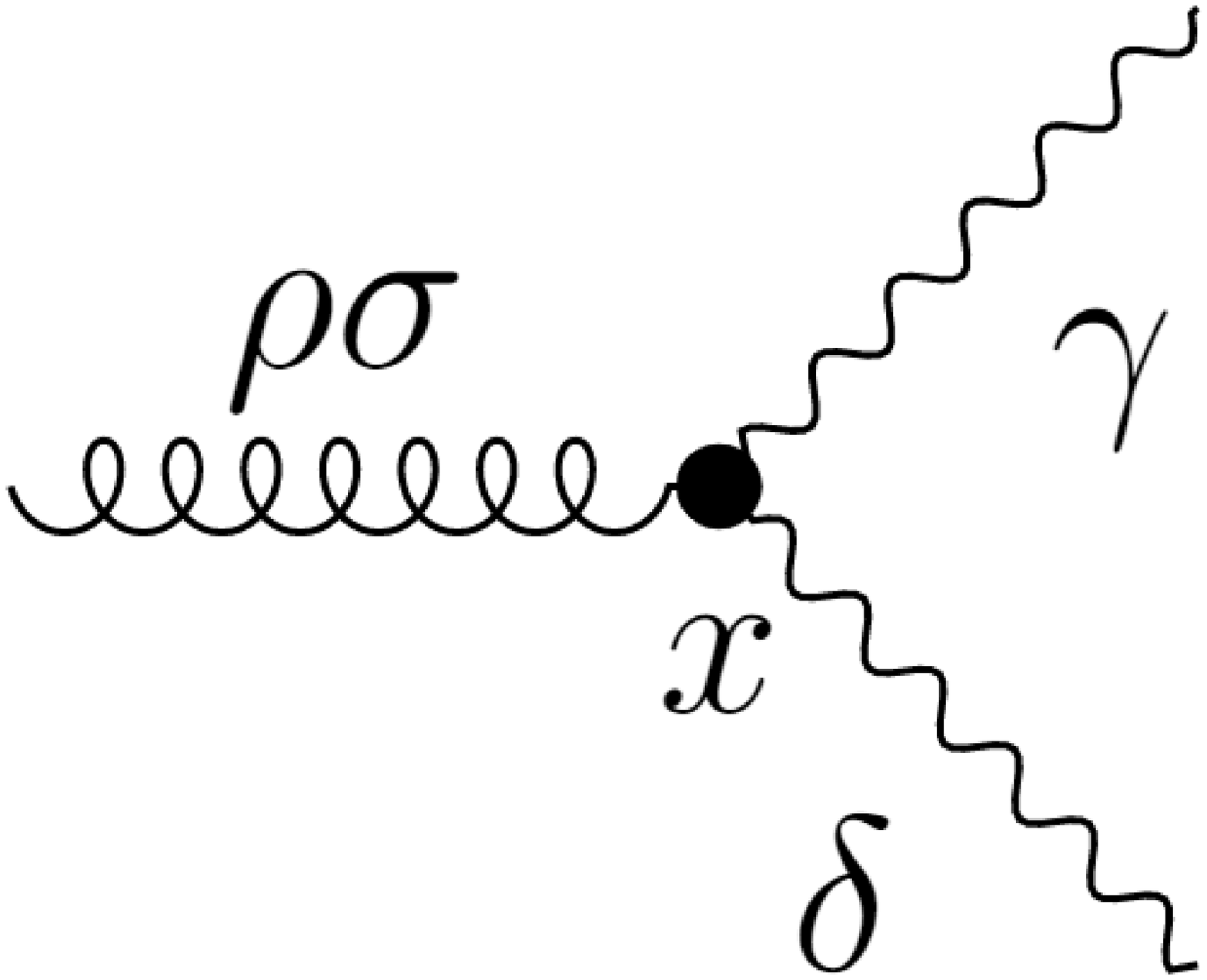}} & \parbox[l]{8.4cm}{\begin{equation}\label{vert_2p1g}\begin{split}&=-i\kappa V^{\gamma\delta\alpha\tau\rho\sigma} \partial_\alpha\uparrow\partial_\tau\downarrow\\
   &\text{where,}\\
   &\ V^{\gamma\delta\alpha\tau\rho\sigma}= \eta^{\rho\sigma}\eta^{\alpha[\tau}\eta^{\delta]\gamma}+4\eta^{\rho)[\gamma}\eta^{\alpha][\delta}\eta^{\tau](\sigma}\end{split}\end{equation}}
\end{tabular}

\section{Appendix: Propagators}

The massless scalar propagator $i\Delta(x;x')$ obeys the equation,
\begin{equation}\label{eq:equation_massless}
    \partial^2 i \Delta(x;x') = i \delta^D(x \!-\! x') \; .
\end{equation}
Even in $D$ spacetime dimensions it has a simple expression in terms of Lorentz 
interval $\Delta x^2(x;x')$,
\begin{equation}\label{massless_scalar_prop}
    i\Delta(x;x') = \frac{\Gamma(\frac{D}{2}-1)}{4\pi^{D/2}}
    \left(\frac{1}{\Delta x^2}\right)^{\frac{D}2-1} \; , 
\end{equation}
where we define,
\begin{equation}
    \Delta x^2(x;x') \equiv ||\Vec{x}-\Vec{x}'||^2 - \left(|t-t'|-i\varepsilon\right)^2 \; .
\end{equation}
The massive scalar propagator $i\Delta_m(x;x')$ obeys the equation,
\begin{equation}\label{massive_scalar_prop}
    (\partial^2 - m^2) i \Delta_m(x;x') = i \delta^D(x \!-\! x') \; .
\end{equation}
It can be written in terms of Bessel functions but the expression itself is not 
necessary for purposes. It turns out that we can always eliminate $i\Delta_{m}(x;x')$,
either with the propagator equation (\ref{massive_scalar_prop}) or by recourse
to one of the Donoghue Identities given in Appendix \ref{sec:donoghue}.

The photon field also requires gauge fixing. The most general Poincar\'e invariant 
gauge fixing functional depends upon an arbitrary parameter $c$,
\begin{equation}
    \mathcal{L}_{\text{EMfix}} = -\frac{1}{2c} (\partial^\mu A_\mu)^2 \; .
\end{equation}
The associated propagator can be expressed using the massless scalar propagator 
(\ref{massless_scalar_prop}),
\begin{equation}
    i[_\rho\Delta_\sigma](x;x') = \left[ \eta_{\rho\sigma} + (c\!-\!1) 
    \frac{\partial_{\rho} \partial_{\sigma}}{\partial^2} \right] i\Delta(x;x') \; .
\end{equation}
The longitudinal term proportional to $c-1$ presumably drops out due to current
conservation but we shall simply adopt the $c=1$ Feynman gauge that Bjerrum-Bohr 
employed \cite{Bjerrum-Bohr:2002aqa},
\begin{equation}
    i[_\rho\Delta_\sigma](x;x') = \eta_{\rho\sigma} i \Delta(x;x') \; .
\end{equation}

The most general Poincar\'e invariant gauge fixing function (\ref{gauge}) depends 
on two parameters $a$ and $b \neq 2$ (for $b = 2$ the gauge fixing functional 
degenerates to the square of a linearized Ricci scalar). To simplify the analysis
we work only to first order in the perturbations $a = 1 + \delta a$ and $b = 1 + 
\delta b$,
\begin{equation}\label{eq:graviton_prop}
\begin{split}
    i[_{\mu\nu}\Delta_{\rho\sigma}](x;x') = \Biggl[ 2 \eta_{\mu(\rho} \eta_{\sigma)\nu}
    -\frac{2 \eta_{\mu\nu} \eta_{\rho\sigma}}{D \!-\! 2} & + \frac{4 \delta a 
    \partial_{(\mu}\eta_{\nu)(\rho}\partial_{\sigma)}}{\partial^2} \\
    & -2 \delta b \left(\eta_{\mu\nu} \frac{\partial_{\rho} \partial_{\sigma}}{\partial^2}
    + \eta_{\rho\sigma} \frac{\partial_{\mu}\partial_{\nu}}{\partial^2} \right)\Biggr]
    i\Delta(x;x') \; .
    \end{split}
\end{equation}

\section{Appendix: The Donoghue Identities}\label{sec:donoghue}

What we term the ``Donoghue Identities'' are not equalities but rather relations
for extracting the nonlocal and nonanalytic parts of amplitudes which can contribute
to long range forces. As originally derived by Donoghue and collaborators
\cite{Donoghue:1993eb,Donoghue:1994dn,Donoghue:1996mt}, they included nonlinear
classical effects as well as quantum effects, but we have retained only the parts
relevant for quantum effects. When expressed in position space these relations all
have the effect of degenerating massive propagators to delta functions. We required
six such relations, of which the final two (those involving factors of $1/\partial^2$) 
were derived by us for this project:
 \begin{itemize}
        \item This concerns 3-point diagrams with no derivatives acting on propagators,
        \begin{equation}\label{eq:3pt}
            i\Delta_m(x;y) i\Delta(x;x') i\Delta(y;x') \longrightarrow 
            \frac{i \delta^D(x \!-\!y)}{2m^2} [i\Delta(x;x')]^2 \; .
        \end{equation}
        \item This concerns 3-point diagrams with a derivative acting on a massless 
        propagator,
        \begin{equation}\label{eq:3pt_derivative}
            \left[\partial_x^{\mu} i\Delta(x;x') \right] i\Delta_m(x;y)
            i\Delta(y;x') \longrightarrow -\partial_{x}^{\mu} 
            \left[\frac{i\delta^D(x \!-\! y)}{2m^2} [i\Delta(x;x')]^2\right] \; .
        \end{equation}
        \item This concerns 3-point diagrams with two derivatives acting on a 
        massless propagator,
        \begin{equation}\label{eq:3pt_2derivative}
            \begin{split}
                \left[\partial_{x}^{\mu} \partial_{x}^{\nu} i\Delta(x;x') \right]
                i\Delta_m(x;y) & \Delta(y;x') \longrightarrow \Bigl\{\partial_{x}^{\mu}
                \partial_{x}^{\nu} \frac{(\partial_{x} \!+\! \partial_{y})^2}{2m^2}
                -\frac{1}{2} \Bigl(\partial_{x}^{\mu} (\partial_{x} \!+\! \partial_{y})^{\nu} \\
                & + \partial_{x}^{\nu} (\partial_{x} \!+\! \partial_{y})^{\mu} \Bigr)
                - \frac{1}{4} \eta^{\mu\nu} (\partial_{x} \!+\! \partial_{y})^2\Bigr\}
                \left[\frac{i\delta^D(x \!-\!y)}{2m^2} [i\Delta(x;x')]^2\right] \; .
            \end{split}
        \end{equation}
        \item These concern 4-point diagrams with no derivatives acting on the propagators. 
        The first is relevant to the box diagrams as shown on the upper part of Figure~\ref{fig:box}. 
        The second is relevant to the cross diagrams as shown on the lower part of Figure~\ref{fig:box},
        \begin{equation}\label{eq:box_dono}
            \begin{split}
                m^2 (\partial_{x} \!+\! \partial_{y})^2 & \left[i\Delta_m(x;y) i\Delta(y;y')
                i\Delta_m(y';x') i\Delta(x';x)\right] \\
                & \longrightarrow \left(1-\frac{\partial_{x} \!\cdot\! \partial_{x'} \!-\! m^2}{3m^2}
                \right)[i\Delta(x;x')]^2 \delta^D(x\!-\!y) \delta^D(x' \!-\!y') \; , \\
                 m^2 (\partial_{x} \!+\! \partial_{y})^2 & \left[i\Delta_m(x;y) i\Delta(y;x')
                 i\Delta_m(x';y') i\Delta(y';x)\right] \\
                 & \longrightarrow \left(-1 \!+\! \frac{\partial_{x}\!\cdot\! \partial_{y'} \!-\! m^2}{3m^2}
                 \right)[i\Delta(x;x')]^2 \delta^D(x\!-\!y) \delta^D(x'\!-\!y') \; .
            \end{split}
        \end{equation}
        \item This concerns 3-point diagrams with a derivative and an inverse Laplacian 
        on a massless propagator,
        \begin{equation}\label{eq:new_dono_1}
            \begin{split}
                i\Delta_m(x;y) & i\Delta(x;x') \frac{\partial^\mu}{\partial^2} \Delta(y;x') \\
                & \longrightarrow \frac{1}{m^2} \Bigl[\frac{(\partial_{x} \!+\! \partial_{y})^{\mu}}{8}
                - \frac{\partial_x^{mu}}{2} + \frac{1}{3m^2} \partial_{x}^{\mu} 
                (\partial_{x} \!+\! \partial_{y})^2 \Bigr] \left[\frac{i\delta^D(x\!-\!y)}{2m^2}
                [i\Delta(x;x')]^2 \right] .
            \end{split}
        \end{equation}
        \item This concerns 3-point diagrams with two derivatives and an inverse Laplacian 
        on a massless propagator,
        \begin{equation}\label{eq:new_dono_2}
            \begin{split}
                i\Delta_m(x;y) i\Delta(x;x')&\frac{\partial^\mu\partial^\nu}{\partial^2}\Delta(y;x')\\
                &\longrightarrow\Bigl[\frac{1}{2}\eta^{\mu\nu}-\frac{1}{m^2}\partial_{x}^\mu\partial_x^\nu+\frac{1}{6m^2}\eta_{\mu\nu}(\partial_x+\partial_y)^2+\frac{1}{2m^2}\Bigl(\partial_{x}^\mu(\partial_x+\partial_y)^\nu\\
                &+\partial_{x}^\nu(\partial_x+\partial_y)^\mu\Bigr)-\frac{2}{3m^4}(\partial_x+\partial_y)^2\partial_x^\mu\partial_x^\nu\Bigr]\left[\frac{i\delta^D(x\!-\!y)}{2m^2}[i\Delta(x;x')]^2\right] .
            \end{split}
        \end{equation}
    \end{itemize}


\begin{thebibliography}{99}

\bibitem{Jackson:1998nia}
J.~D.~Jackson,
``Classical Electrodynamics,'' (Wiley \& Sons, New York, 1998).

\bibitem{Schwinger:1960qe}
J.~S.~Schwinger,
J. Math. Phys. \textbf{2}, 407-432 (1961)
doi:10.1063/1.1703727

\bibitem{Mahanthappa:1962ex}
K.~T.~Mahanthappa,
Phys. Rev. \textbf{126}, 329-340 (1962)
doi:10.1103/PhysRev.126.329

\bibitem{Bakshi:1962dv}
P.~M.~Bakshi and K.~T.~Mahanthappa,
J. Math. Phys. \textbf{4}, 1-11 (1963)
doi:10.1063/1.1703883

\bibitem{Bakshi:1963bn}
P.~M.~Bakshi and K.~T.~Mahanthappa,
J. Math. Phys. \textbf{4}, 12-16 (1963)
doi:10.1063/1.1703879

\bibitem{Keldysh:1964ud}
L.~V.~Keldysh,
Zh. Eksp. Teor. Fiz. \textbf{47}, 1515-1527 (1964)

\bibitem{Chou:1984es}
K.~c.~Chou, Z.~b.~Su, B.~l.~Hao and L.~Yu,
Phys. Rept. \textbf{118}, 1-131 (1985)
doi:10.1016/0370-1573(85)90136-X

\bibitem{Jordan:1986ug}
R.~D.~Jordan,
Phys. Rev. D \textbf{33}, 444-454 (1986)
doi:10.1103/PhysRevD.33.444

\bibitem{Calzetta:1986ey}
E.~Calzetta and B.~L.~Hu,
Phys. Rev. D \textbf{35}, 495 (1987)
doi:10.1103/PhysRevD.35.495

\bibitem{Ford:2004wc}
L.~H.~Ford and R.~P.~Woodard,
Class. Quant. Grav. \textbf{22}, 1637-1647 (2005)
doi:10.1088/0264-9381/22/9/011
[arXiv:gr-qc/0411003 [gr-qc]].

\bibitem{Leonard:2012fs}
K.~E.~Leonard and R.~P.~Woodard,
Phys. Rev. D \textbf{85}, 104048 (2012)
doi:10.1103/PhysRevD.85.104048
[arXiv:1202.5800 [gr-qc]].

\bibitem{Donoghue:1993eb}
J.~F.~Donoghue,
Phys. Rev. Lett. \textbf{72}, 2996-2999 (1994)
doi:10.1103/PhysRevLett.72.2996
[arXiv:gr-qc/9310024 [gr-qc]].

\bibitem{Donoghue:1994dn}
J.~F.~Donoghue,
Phys. Rev. D \textbf{50}, 3874-3888 (1994)
doi:10.1103/PhysRevD.50.3874
[arXiv:gr-qc/9405057 [gr-qc]].

\bibitem{Bjerrum-Bohr:2002fji}
N.~E.~J.~Bjerrum-Bohr, J.~F.~Donoghue and B.~R.~Holstein,
Phys. Rev. D \textbf{68}, 084005 (2003)
[erratum: Phys. Rev. D \textbf{71}, 069904 (2005)]
doi:10.1103/PhysRevD.68.084005
[arXiv:hep-th/0211071 [hep-th]].

\bibitem{Bjerrum-Bohr:2002gqz}
N.~E.~J.~Bjerrum-Bohr, J.~F.~Donoghue and B.~R.~Holstein,
Phys. Rev. D \textbf{67}, 084033 (2003)
[erratum: Phys. Rev. D \textbf{71}, 069903 (2005)]
doi:10.1103/PhysRevD.71.069903
[arXiv:hep-th/0211072 [hep-th]].

\bibitem{Bjerrum-Bohr:2002aqa}
N.~E.~J.~Bjerrum-Bohr,
Phys. Rev. D \textbf{66}, 084023 (2002)
doi:10.1103/PhysRevD.66.084023
[arXiv:hep-th/0206236 [hep-th]].

\bibitem{Miao:2017feh}
S.~P.~Miao, T.~Prokopec and R.~P.~Woodard,
Phys. Rev. D \textbf{96}, no.10, 104029 (2017)
doi:10.1103/PhysRevD.96.104029
[arXiv:1708.06239 [gr-qc]].

\bibitem{Donoghue:1996mt}
J.~F.~Donoghue and T.~Torma,
Phys. Rev. D \textbf{54}, 4963-4972 (1996)
doi:10.1103/PhysRevD.54.4963
[arXiv:hep-th/9602121 [hep-th]].

\bibitem{Deser:1957zz}
S.~Deser,
Rev. Mod. Phys. \textbf{29}, 417 (1957)
doi:10.1103/RevModPhys.29.417

\bibitem{DeWitt:1960fc}
B.~S.~DeWitt and R.~W.~Brehme,
Annals Phys. \textbf{9}, 220-259 (1960)
doi:10.1016/0003-4916(60)90030-0

\bibitem{Tsamis:1992xa}
N.~C.~Tsamis and R.~P.~Woodard,
Commun. Math. Phys. \textbf{162}, 217-248 (1994)
doi:10.1007/BF02102015

\bibitem{Woodard:2004ut}
R.~P.~Woodard,
[arXiv:gr-qc/0408002 [gr-qc]].

\bibitem{Leonard:2013xsa}
K.~E.~Leonard and R.~P.~Woodard,
Class. Quant. Grav. \textbf{31}, 015010 (2014)
doi:10.1088/0264-9381/31/1/015010
[arXiv:1304.7265 [gr-qc]].

\bibitem{Glavan:2013jca}
D.~Glavan, S.~P.~Miao, T.~Prokopec and R.~P.~Woodard,
Class. Quant. Grav. \textbf{31}, 175002 (2014)
doi:10.1088/0264-9381/31/17/175002
[arXiv:1308.3453 [gr-qc]].

\bibitem{Wang:2014tza}
C.~L.~Wang and R.~P.~Woodard,
Phys. Rev. D \textbf{91}, no.12, 124054 (2015)
doi:10.1103/PhysRevD.91.124054
[arXiv:1408.1448 [gr-qc]].

\bibitem{Glavan:2015ura}
D.~Glavan, S.~P.~Miao, T.~Prokopec and R.~P.~Woodard,
Class. Quant. Grav. \textbf{32}, no.19, 195014 (2015)
doi:10.1088/0264-9381/32/19/195014
[arXiv:1504.00894 [gr-qc]].

\bibitem{Glavan:2016bvp}
D.~Glavan, S.~P.~Miao, T.~Prokopec and R.~P.~Woodard,
Class. Quant. Grav. \textbf{34}, no.8, 085002 (2017)
doi:10.1088/1361-6382/aa61da
[arXiv:1609.00386 [gr-qc]].

\bibitem{Glavan:2019msf}
D.~Glavan, S.~P.~Miao, T.~Prokopec and R.~P.~Woodard,
JHEP \textbf{10}, 096 (2019)
doi:10.1007/JHEP10(2019)096
[arXiv:1908.06064 [gr-qc]].

\end{thebibliography}
\end{document}